\documentclass{aa}  
\usepackage{booktabs,threeparttable}
\usepackage{balance}
\usepackage{graphicx}
\usepackage{xcolor}
\usepackage{txfonts}

\usepackage{array,multirow}
\newcolumntype{C}[1]{>{\centering\arraybackslash}m{#1}}
\newcolumntype{L}[1]{>{\raggedright\arraybackslash}m{#1}}

\usepackage[normalem]{ulem}

\mathchardef\mhyphen="2D

\newcommand{\civ}{\ifmmode {\rm C}\,{\sc iv} \else C\,{\sc iv}\fi}
\newcommand{\CIV}{\ifmmode {\rm C}\,{\sc iv}\,\lambda1549 \else 
	           C\,{\sc iv}\,$\lambda1549$\fi}
\newcommand{\oiii}{O\,{\sc iii}}

\newcommand{\Lo}{L_{\rm UV}}

\newcommand{\Lx}{L_{\rm X}}

\newcommand{\chandra}{\emph{Chandra}}

\begin{document}

   \title{The \emph{Chandra} view of the relation between X-ray and UV emission in quasars\thanks{Table 2 is only available in electronic form
at the CDS via anonymous ftp to cdsarc.u-strasbg.fr (130.79.128.5)
or via http://cdsweb.u-strasbg.fr/cgi-bin/qcat?J/A+A/}}
\titlerunning{The $\Lx-\Lo$ relation in the SDSS - CSC 2.0 and \emph{Chandra} COSMOS Legacy quasar sample}

   \author{S. Bisogni
          \inst{1}\fnmsep \inst{2} \fnmsep\thanks{susanna.bisogni@inaf.it}
          \and
          E. Lusso
          \inst{3}\fnmsep \inst{4}
          \and
          F. Civano
          \inst{2}
          \and
          E. Nardini 
          \inst{3}\fnmsep \inst{4}
          \and
          G. Risaliti
          \inst{3}\fnmsep \inst{4}
          \and
          M. Elvis
          \inst{2}
          \and
          G. Fabbiano
          \inst{2} 
          }

   \institute{INAF – Istituto di Astrofisica Spaziale e Fisica Cosmica Milano, Via Corti 12, 20133 Milano, Italy
         \and
             Harvard-Smithsonian Center for Astrophysics, Cambridge, MA 02138, USA
         \and
            Dipartimento di Fisica e Astronomia, Universit\`a di Firenze, via G. Sansone 1, 50019 Sesto Fiorentino, Firenze, Italy
         \and
            INAF – Osservatorio Astrofisico di Arcetri, 50125 Florence, Italy
             }


  \abstract
{We present a study of the relation between X-rays and ultraviolet emission in quasars for a sample of broad-line, radio-quiet objects obtained from the cross-match of the Sloan Digital Sky Survey DR14 with the latest \emph{Chandra} Source Catalog 2.0 (2,332 quasars) and the \emph{Chandra} COSMOS Legacy survey (273 quasars). 
The non-linear relation between the ultraviolet (at 2500\,\AA, $\Lo$) and the X-ray (at 2 keV, $\Lx$) emission in quasars has been proved to be characterised by a smaller intrinsic dispersion than the observed one, as long as a homogeneous selection, aimed at preventing the inclusion of contaminants in the sample, is fulfilled. 
By leveraging on the low background of \emph{Chandra}, we performed a complete spectral analysis of all the data available for the SDSS-CSC2.0 quasar sample (i.e. 3,430 X-ray observations), with the main goal of reducing the uncertainties on the source properties (e.g. flux, spectral slope).
We analysed whether any evolution of the $\Lx-\Lo$ relation exists by dividing the sample in narrow redshift intervals across the redshift range spanned by our sample, $z \simeq 0.5-4$. We find that the slope of the relation does not evolve with redshift and it is consistent with the literature value of $0.6$ over the explored redshift range, implying that the mechanism underlying the coupling of the accretion disc and hot corona is the same at the different cosmic epochs. We also find that the dispersion decreases when examining the highest redshifts, where only pointed observations are available. These results further confirm that quasars are `standardisable candles', that is we can reliably measure cosmological distances at high redshifts where very few cosmological probes are available.}

               
   \maketitle


\section{Introduction} \label{sec:intro}
Although the non-linear relation of ultraviolet ($L_{\rm UV}$) and X-ray ($L_{\rm X}$) luminosities in quasars is well known \citep{Tananbaum1979, Zamorani1981,AvniTananbaum1986,Wilkes1994,Steffen2006,Just2007,Young2010,Lusso2010}, we still lack the understanding of the physical mechanism connecting the accretion disc and the hot corona. Photons from the accretion disc (emitting in the UV) are scattered to higher energies (X-rays) by the relativistic electrons present in a hot plasma (the so-called corona) through inverse Compton \citep[e.g.][]{HaardtMaraschi1991,HaardtMaraschi1993}. Nonetheless, for the hot corona not to cool down very fast, there must be a mechanism transferring part of the gravitational energy lost during the accretion from the disc to the corona \citep[e.g.][]{MerloniFabian2001}, which likely involves magnetic reconnection \citep{DiMatteo1998, Liu2002}, but whose details are hardly known. 

An increase by a factor of 10 in the UV emission corresponds to an increase by only a factor of 4 at X-ray energies, meaning that more luminous quasars in the UV are relatively less luminous in the X-rays. This effect is parametrised as a non-linear relation, which is usually expressed in terms of the logarithms:
\begin{equation}
\log(\Lx) = \gamma \log(\Lo) + \beta \,,
\end{equation}
where the proxies used for the UV and X-ray emissions are the (rest-frame) $2500$ \AA\ and 2 keV monochromatic luminosities, respectively, and the slope has been found to be $\gamma \simeq 0.6 \pm 0.1$ in all the works mentioned above.
This relation holds for several decades in both UV and X-ray luminosity \citep[e.g.][]{Steffen2006,Lusso2010,Salvestrini2019} and it does not appear to evolve with redshift \citep[e.g.][]{Vignali2003, Vagnetti2010}, suggesting that the physical mechanism behind the disc-corona synergy must be universal, namely the same from the faintest to the brightest objects of the active galactic nuclei (AGN) population and at different ages of the Universe. Still, we have very few clues on what the physical process at work is.

Our group has investigated the $\Lx-\Lo$ relation, with the twofold aim of understanding its physical meaning \citep{LussoRisaliti2016,LussoRisaliti2017} and testing its application for cosmological purposes \citep{RisalitiLusso2015,RisalitiLusso2019,Bisogni2018,Lusso2019,lusso2020,bargiacchi2021}. 
Thanks to the non-linearity of this relation, we can have an independent measurement of the AGN cosmological distances once we convert luminosities into fluxes as:
\begin{equation}
\log (f_{\rm X}) = \gamma \, \log (f_{\rm UV}) + (2-2\gamma) D_{\rm L} + \beta'\,, 
\end{equation}
where the luminosity distance $D_{\rm L}$ is not canceled out if $\gamma\neq 1$ and $\beta'$ depends on slope and intercept as $\beta' = \beta + (\gamma-1)\rm{log}4\pi$.
On these grounds, it is then possible to build a Hubble diagram for quasars \citep{RisalitiLusso2015,RisalitiLusso2019} and use these objects as cosmological probes, bridging the gap from the farthest observed supernovae Ia (at $z \simeq 2$, \emph{Pantheon} sample; \citealt{Scolnic2018}) to considerably higher redshifts ($z\simeq 7.5$, \citealt{Banados2018}).

In both the physical and the cosmological applications, the analysis of the dispersion in the $\Lx-\Lo$ relation is crucial. In the first case, it can constrain the coupling of the AGN disc and corona. In the second, it determines the precision achievable in the measure of cosmological distances, that is in the extraction of cosmological information.

A substantial progress came with 
the discovery that most of the observed dispersion, $\delta_{obs}$, is not intrinsic \citep{LussoRisaliti2016,LussoRisaliti2017}. When the contributions to the dispersion ascribable to observational issues can be properly removed - or at least reduced - $\delta_{obs}$ significantly decreases (see also \citealt{lusso2020}). 
Several contributors to the dispersion have been found, such as 1.) dust reddening and X-ray absorption, which can prevent the measure of the intrinsic fluxes at 2500\,\AA\ and 2 keV, respectively; 2.) calibration issues, especially in the X-ray band, related to the serendipitous detection of many of the sources in the X-ray catalogues, with a range of distances from the aim-point of the observation; 3.) the inclusion in the sample of sources detected following a positive fluctuation with respect to their average emission (Eddington bias); 4.) the inclination of the accretion disc with respect the line of sight, affecting the measure of the intrinsic UV flux; 5.) emission variability in both bands; 6.) non-simultaneity of the UV and X-ray emission measurements. Some of these factors are in principle correctable (1--5), while we can only be aware of others (e.g. 6).

In a recent work, \cite{RisalitiLusso2019} selected $\sim$1,600 quasars from a parent sample of $\sim$8,000 objects, obtained through the cross-match of the SDSS DR7+DR12 and 3XMM DR7. By excluding from the sample 
the quasars affected by the observational problems listed above, and by using, mainly, X-ray photometric data, they found a dispersion $\delta = 0.24$ dex, to be compared to the initial value of $\delta\sim 0.40$ dex of the parent sample. The latter value is consistent with the one found since the earliest works, which had deterred, until very recently, from using the $\Lx-\Lo$ relation for any cosmological application. Such a reduction in the dispersion allowed the detection, to a $\sim$3-4$\sigma$ significance level, of a discrepancy between the cosmological parameters measured at $z<1.4$ and at $z>1.4$, which is compatible with the tension in the value of the Hubble constant $H_{0}$ measured with the Distance Ladder methods \citep{Riess2019} and that measured from the temperature fluctuations of the Cosmic Microwave Background assuming the concordance $\Lambda$CMD model \citep{PlanckCollaboration2018}.

In the wake of these achievements, here we present the analysis of a brand new sample of 7,036 SDSS DR14--\emph{Chandra} Source Catalog 2.0 quasars. The main goals of this work are 1.) exploiting the low background level of \emph{Chandra} to further decrease the observed dispersion in the $\Lx-\Lo$ relation by measuring the 2 keV fluxes spectroscopically; 2.) examining the \emph{non}-evolution of the slope of the relation with redshift, crucial to both its employment in cosmological applications and the understanding of 
quasar evolution across cosmic time; and 3.) accounting for the contribution to the dispersion of the X-ray variability through the use of multiple observations of the sources in the sample.

This paper is organised as follows: in Section~\ref{sec:sample_intro} we present the sample and the preliminary selection criteria. In Section~\ref{sec:data_analysis} we describe how the fluxes in the UV and X-ray bands are retrieved. For the X-ray band, the procedure used to infer the flux limit for each observation and the criteria for assuring that no Eddington bias is present in the sample are presented. In Section~\ref{sec:analysis} we describe the analysis on the $f_{\rm X}-f_{\rm UV}$ relation and its 
possible evolution with redshift, and the analysis of the $\Lx-\Lo$ relation for the total sample. We present our results and conclusions in Section~\ref{sec:discussion} and Section~\ref{sec:conclusions}, respectively. 
Throughout the paper, when computing luminosities, we assume a concordance flat $\Lambda$CDM model, with $H_{0} = 70$ km s$^{-1}$ Mpc$^{-1}$, $\Omega_{M}=0.3$ and $\Omega_{\Lambda}=1-\Omega_{M}$.


\section{Data and sample selection} \label{sec:sample_intro}

Our parent sample includes sources selected from the following catalogues:

\begin{itemize}

\item[1.]  The cross-match of the fourteenth data release of the {\it Sloan Digital Sky Survey quasar catalogue}\footnote{\url{https://www.sdss.org/dr14/algorithms/qso_catalog/}} \citep{Paris2018DR14} and the latest release of the {\it Chandra Source Catalog} \citep[CSC 2.0\footnote{\url{http://cxc.harvard.edu/csc/}};][]{Evans2010}. The SDSS DR14 quasar catalogue includes the previous releases (DR7 and DR12) and expands them with newly observed sources, for a total of more than 500,000 spectroscopically confirmed quasars. 
The CSC 2.0 contains information on $\sim$315,000 X-ray sources detected in 10,382 \chandra\ ACIS and HRC-I imaging observations publicly released prior of 2015.
We obtained 7,036 unique matches within 3$\arcsec$ from the SDSS DR14 coordinates, corresponding to more than 10,000 single \chandra\ observations in the X-ray band, when multiple observations of the same source are taken into account.

\item[2.] Unobscured (type-1) AGN from the {\it Chandra COSMOS Legacy Survey} \citep{Civano2016}, a 4.6-Ms \emph{Chandra} programme providing a coverage of $\sim$160 ks for the central 1.5 deg$^{2}$ and of $\sim$80 ks for the outer region of the 2.2 deg$^{2}$ COSMOS field\footnote{This programme is an extension of the previous \emph{Chandra} COSMOS survey \citep[C-COSMOS,][]{Elvis2009, Puccetti2009, Civano2012}, which already covered with 1.8 Ms of \emph{Chandra} observations $\sim$1/4 of the COSMOS field at the same depth plus 0.5 deg$^{2}$ at $\sim$80 ks depth.}.
We selected type-1 AGN following the spectroscopic and photometric identification presented in \cite{Marchesi2016}. As a first requirement, we asked for secure associations with the near-infrared counterparts: this selected 3,910 of the initial 4,016 sources. When a spectroscopic observation is available, the sources are identified as Broad Line AGN (BLAGN) if they present at least one broad line with FWHM $> 2000$ km/s (635 type-1 AGN out of the total 2,262/3,910 with a spectrum), whilst 3,798/3,910 sources have a spectral energy distribution (SED) fitting classification (880/3,798 SEDs are reproduced by a type-1 AGN template in the optical/UV). When a spectrum is available, the spectroscopic classification is preferred over the one obtained from the SED analysis. 
Our selection yielded a sample of 1,001 type-1 AGN (635 spectroscopically identified and 366 classified only through SED fitting), $\sim$25\% of the whole \emph{Chandra COSMOS Legacy} catalogue, with an unambiguous match between X-ray detection and NIR counterpart and either a spectral or a photometric classification.
Only 943 out of the 1,001 X-ray selected type-1 objects have a detection in the Ultra-VISTA survey \citep{McCracken2012} and the required photometric information to perform the analysis of their SED and to infer their $2500$~\AA\ flux. 
Quasars in the COSMOS sample that overlapped with the SDSS-CSC2.0 one were excluded from the final COSMOS sample, yielding $882$ objects.
Throughout this work we made use of the photometric information listed in the Near Infrared selected catalogue COSMOS2015 \citep{Laigle2016}. 
\end{itemize}

\subsection{Sample selection in the optical/UV band} \label{sec:sample}

\begin{table}[ht]
\footnotesize
\caption{Summary of the observations.}
\begin{center}
\renewcommand{\arraystretch}{1.0}
\begin{tabular*}{1.0\linewidth}{@{\extracolsep{\fill}}l c c c c}
\hline 

 \multicolumn{5}{c}{{\bfseries SDSS - CSC 2.0}}  \\

\hline 
UV pre-selection                                  &          DR7         &        DR12        &        DR14    &  TOT    \\
                                                  &                         &                         &                   &    sources       \\
\hline
SDSS vs CSC 2.0 (3$\arcsec$)          &  $2348$            &     $3276$        &      $1412$        &       $7036$  \\
non-BAL (flags)            &   $2217$            &     $3111$        &      $1412$        &       $6740$   \\
non-BAL (Gibson09)    &   $2211$            &     $3111$        &      $1412$       &       $6734$    \\
non-BAL (Allen11)      &   $2206$            &     $3111$        &      $1412$       &       $6731$    \\
non-RL (flags)            &   $1922$            &     $3111$        &      $1412$       &       $6445$    \\
non-RL (Mingo16 MIXR)   &   $1919$            &     $3096$        &      $1412$       &       $6427$    \\
non-RL (Mingo16 crit.)  &   $1919$            &     $3011$        &      $1376$       &       $6306$    \\
dust-free                  &   $1729$            &     $1993$        &      $1027$        &       $4749$   \\
host-free     &   $1600$            &     $1965$        &      $1013$        &       $4578$   \\

\hline
X-ray pre-selection                            &                        &                        &                    &  TOT    \\
                                                  &                         &                         &                   &    observations      \\
                                                  &                         &                         &                   &    (sources)       \\
\hline
SDSS vs CSC 2.0 obs  &   $2294$             &     $3318$        &      $1265$      &       $6877$  \\
                                 &   $(1559)$          &    $(1911)$        &    $(976)$      &    $(4446)$ \\

soft band   &   $1988$             &     $2636$        &      $1000$  &       $5624$  \\
                 &  $(1189)$            &    $(1141)$       &    $(567)$   &      $(2897)$  \\

$\Theta < 10$ arcmin  &   $1510$       &     $1998$       &      $704$     &       $4212$  \\
                                   &  $(975)$        &    $(1014)$      &   $(477)$      &    $(2466)$ \\

available spectra   &   $1362$         &     $1642$    &      $565$       &       $3569$  \\
                           &  $(922)$          &    $(995)$     &     $(475)$     &      $(2392)$ \\

flux limit  &   $1322$         &     $1560$    &      $548$         &       $3430$  \\
               &   $(891)$        &    $(967)$     &     $(474)$       &      $(2332)$ \\

\hline
X-ray selection                            &                        &                        &                    &  TOT    \\
                                                  &                         &                         &                   &    observations      \\
                                                  &                         &                         &                   &    (sources)       \\
\hline
$\Gamma$ &   $998$         &     $1066$    &      $378$         &       $2442$  \\
               &   $(683)$        &    $(697)$     &     $(337)$       &      $(1717)$  \\

Eddington bias   &   $706$         &     $535$    &      $144$         &       $1385$  \\
                        &   $(477)$        &    $(350)$     &     $(131)$       &      $(958)$ \\
\hline

 \multicolumn{5}{c}{{\bfseries Chandra COSMOS Legacy}}  \\
\hline 
UV pre-selection              & & &         &    TOT                        \\
                                       & & &                       &    sources                  \\
\hline
COSMOS Type-1 AGN     &  & & &     $1001$            \\
ultra-vista data                                      &  & & &     $882$           \\
non-BAL (flags)                                      &    & & &   $875$           \\
non-BAL (Gibson09)                                &    & & &   $875$           \\
non-BAL (Allen11)                                    &  & & &     $875$           \\
non-RL (flags)                                         &    & & &   $872$           \\
non-RL ($R>10$)                                        &   & & &    $823$           \\
dust-free                                                 &  & & &     $289$           \\
host-free                                                 &  & & &     $288$           \\
\hline
X-ray pre-selection                            &   & & &      TOT                  \\
                                                       &   & & &       sources          \\
\hline                                                   
soft band                                          &   & & &    $273$           \\
\hline

X-ray selection                            &                        &                        &                    &  TOT    \\
                                                  &                         &                         &                   &    sources       \\
\hline
$\Gamma$ &                &           &               &       $151$  \\

Eddington bias   &             &            &               &       $140$  \\

\hline

\end{tabular*}
{\raggedright {\bfseries Notes.} Summary of the observations statistics among the different sub-samples and categories. Numbers in parentheses are sources corresponding to observations. \par}
\label{tab1}
\end{center}
\end{table}
For the sample selection, we followed the approach described in \cite{lusso2020}. Briefly, we excluded from the sample broad absorption line (BAL) and radio-loud (RL) quasars. The presence of obscuration \citep{Murray1995, Elvis2000} or an additional contribution to the X-ray emission due to winds and jets \citep[e.g.][]{Wilkes1987} require a complex modelling of the AGN spectra to disentangle these processes from the intrinsic emission of the X-ray corona, which in turn increases the uncertainty on the X-ray flux measurement. We finally selected sources that are not dust-absorbed or host-galaxy contaminated in the UV/optical band.
\subsubsection{Broad absorption line quasars}
\noindent{\it {\bfseries SDSS-CSC2.0.}} To exclude BALs, we used the DR7 and DR12 dedicated flags provided by the quasar catalogue published by \cite{Shen2011} and the visual BAL flag provided by \cite{Paris2017DR12}, respectively. To exclude as many BALs as possible among sources observed for the first time in DR14, we used the Balnicity Index (BI) as defined in \cite{Weymann1991} and provided by \cite{Paris2018DR14}. Since BAL quasars are traditionally defined as having BI(\civ)\,$>0$, we required BI(\civ)\,$=0$.
The exclusion of sources through the BI is not as accurate as a visual inspection, but it assures to not exclude non-BAL objects from the sample. With this selection we obtained $6,740$ objects. We then cross-matched the sample with the BAL sample by \cite{Gibson2009} and with the one by \cite{Allen2011}, finding 6 and 5 more BALs, respectively. As discussed in \cite{Paris2018DR14}, these two works adopt slightly different fitting strategies in the evaluation of quasar emission, implying differences in the BAL distributions of the samples. Some objects can therefore be classified as BAL by \cite{Gibson2009} and/or \cite{Allen2011}, but not as such in \cite{Paris2018DR14}. To be conservative, we excluded all the objects classified as BAL in at least one of these three works. The final `BAL-free' sample is composed by 6,729  quasars, $96\%$ of the parent sample (Tab.~\ref{tab1}).

\noindent {\it {\bfseries COSMOS.}} For the objects present in either the DR7 (10 sources) or DR12 (19) SDSS quasar catalogues, we used the dedicated flag and exclude 7 BALs. The cross-match with the \cite{Gibson2009} and \cite{Allen2011} did not give any result. The BAL free sample is composed by 875 sources.

\subsubsection{Radio-Loud quasars}
\noindent {\it {\bfseries SDSS-CSC2.0.}} We first excluded 284 DR7 quasars with a value of the radio loudness parameter, $R$, higher than 10 \citep{Kellermann1989}, where $R=f_{\nu, 6 \mathrm{cm}}/f_{\nu, 2500}$ is listed in the \cite{Shen2011} catalogue. We then cross-matched the sample with the MIXR catalogue of 2,753 objects, obtained through the cross-match of the largest catalogues available in the Mid-Infrared (\emph{WISE}), X-ray (3XMM-DR5) and Radio (FIRST/NVSS) and published by \cite{Mingo2016}. This yielded 44 sources within $3\arcsec$ of positional error, 18 of which are classified as RL and hence excluded from our sample.
The analysis of \cite{Mingo2016} on the multi-wavelength MIXR sample shows also that, regardless of their properties in other bands, sources with a $L_{1.4\mathrm{GHz}}>5 \times 10^{41}$ can be classified as RL. We therefore cross-matched our sample with the FIRST/NVSS catalogue and retrieved the rest-frame $L_{1.4\mathrm{GHz}}$ by assuming a radio spectral slope of 0.5 \citep{Hao2014}. Within a maximum positional error of 30$\arcsec$, we found 346 matches, of which 121 are classified as RL following the criterion on $L_{1.4\mathrm{GHz}}$.
The `RL-free' sample consists of 6,306 quasars.

\noindent {\it {\bfseries COSMOS.}} For the sources in the SDSS DR7 quasars catalogue, we used the dedicated flag, excluding 3 RL sources. Once the UV SED analysis was performed (Appendix \ref{sec:appendix1}) and the 4400\,\AA\ luminosity was available, we computed the radio-loudness parameter $R=L_{5 \rm GHz}/L_{4400}>10$, using the \cite{Smolcic2017} catalogue at 3\,GHz to infer the luminosity at 5\,GHz, and excluded 49 more sources, leaving 823 sources in the `RL-free' COSMOS sample.

\subsubsection{Dust reddening and host galaxy contamination}
\noindent {\it {\bfseries SDSS-CSC2.0.}} Dust along the line of sight and contamination by the host galaxy are the main reasons for an incorrect measure of the intrinsic 2500\,\AA\ flux, proxy for the accretion disc emission. In order to keep in the sample only sources with a measure of the 2500\,\AA\ close to the intrinsic one, we adopted the same criteria as in \cite{LussoRisaliti2016}, by computing the spectral slopes of the optical spectra so to evaluate to what degree the continuum is affected by dust absorption or by galaxy contamination (see Appendix \ref{sec:appendix1} and Fig. \ref{fig:reddening_selection} for details). We included in the sample only the sources with all the five SDSS bands magnitudes available (6,282/6,306) and selected 4,749 blue, host-galaxy- and dust-free quasars.
Finally, by choosing only sources with $z>0.48$, we excluded 171 more objects, obtaining a selection of 4,578 quasars.
Sources at lower redshifts are more affected by host-galaxy contamination for two main reasons: first, low-redshift AGN are on average intrinsically fainter than their higher-redshift companions and do not outshine their hosts to the same extent; second, the BOSS spectrograph wavelength coverage (3650--10400\,\AA) sets a lower limit for the 2500\,\AA\ to be inferred without need for extrapolation. The combination of the two effects could lead to an overestimation of the 2500\,\AA\ flux measure for sources at $z<0.48$.

\noindent {\it {\bfseries  COSMOS.}} By applying the same criteria on the spectral slopes computed with the UV SED analysis, we excluded 534 reddened sources. By considering only $z>0.48$ sources, the sample reduced to 288 objects. The \emph{Chandra} COSMOS Legacy sources are X-ray selected, and therefore biased towards a more obscured population of quasars, as it is evident comparing the spectral slopes in the UV/optical/NIR bands for the COSMOS sample with those of the SDSS (optically selected) one (Fig. \ref{fig:reddening_selection}). The fraction of discarded sources through this last selection is large, about 60\% of the initial sample (882 quasars).

\subsection{Sample selection in the X-ray band} \label{sec:Xsample}

\noindent {\it {\bfseries SDSS-CSC2.0.}} Along with tabulated source properties, the CSC 2.0 provides data products - ready to be used for scientific analysis - that can be directly downloaded using the \emph{CSCview} application\footnote{\url{http://cda.harvard.edu/cscview/}}. The precision of the 2 keV flux measurement is one of the major concerns in this study. For this reason, we performed a complete spectral analysis using the X-ray spectra available for all the observations of  X-ray sources and retrievable from the CSC 2.0 archive as data products.

As mentioned above, CSC 2.0  is built on the 10,382 \emph{Chandra} ACIS and HRC-I imaging observations publicly released prior of 2015.
The CSC 2.0 performs source detection on the sum of overlapping observations (stacks), to increase the sensitivity, but includes source properties computed both at the level of the stack and also at the level of the single observations contributing to the stack. 
As we are interested in estimating the contribution of the X-ray variability to the observed dispersion of the relation, including in the sample those sources with multiple observations is a basic requirement. For the 4,446 sources in our sample, we found 6,877 observations (Tab. \ref{tab1}) and the corresponding spectra. 
With a preliminary selection, we restricted the observations to those with: 
\begin{itemize}
\item[1.] a measurement of the flux in the soft (0.5--2 keV) or in both the soft and the hard band (2--7 keV) and 
\item[2.] an off-axis angle (the distance of the source from the aim-point) $<$10$\arcmin$ in that observation, which provides the best of spatial resolution. 
\end{itemize}
Due to the quality of the X-ray measurements for sources that are mainly serendipitous in X-ray catalogues, the largest among the contributions to the dispersion are ascribable to the Eddington bias and to the accuracy and precision of the X-ray fluxes \citep{RisalitiLusso2015, LussoRisaliti2016, LussoRisaliti2017, RisalitiLusso2019}. By using \emph{Chandra} spectra to directly measure the X-ray flux and limiting our selection to on-axis sources, we are partially addressing these issues. 
Given the importance of the determination of the flux limit for each observation - enabling us to discard the sources 
included in the sample because of a positive fluctuation compared to their average intensity  - we kept only those observations
for which such a measure is possible. This selected 3,430 observations (Tab. \ref{tab1}), corresponding to 2,332 sources, of which 439 have multiple (i.e. two or more) observations\footnote{We note that we examined \emph{all} the X-ray observations available for the sources we pre-selected in the UV band. As a result, we analysed 3,430 spectra extracted at the source region from \emph{Chandra} imaging observations, surviving the selection criteria described above in terms of off-axis angle and of availability of flux measurement. The forthcoming X-ray analysis was also carried out at the observation-level, unless otherwise specified.}. The distributions of soft-band signal-to-noise ratio and redshift are shown in Fig. 1 and 2, respectively.

\noindent {\it {\bfseries COSMOS.}} In this case, with a median exposure time of $\sim$156 ($\sim$159) ks in the soft (hard) band, compared to a median effective exposure time of $\sim$20 ks for CSC 2.0 observations, we chose to use the X-ray photometric measurements already available in the \emph{Chandra} COSMOS Legacy catalogue \citep{Civano2016}. Requiring the availability of at least the soft band selected 273 sources, 230 of which also have the hard band available. The flux limit in this survey is quite uniform and is reported in \cite{Civano2016}.\\

\section{Data analysis} \label{sec:data_analysis}
In this section, we report how the monochromatic fluxes have been computed from the multi-wavelength data collected, and how the sample has been corrected for the Eddington bias. 

\subsection{Computation of UV fluxes}
\label{subsec:UV_SED}
The flux density at 2500\,\AA\ for all the sources in the sample was obtained through an analysis of the SED, which made use of the multi-wavelength photometric information from the UV to near-infrared bands, available for both the SDSS-CSC2.0 and the \emph{Chandra} COSMOS Legacy samples. The determination of SED slopes allowed the exclusion of the dust-affected sources in the sample, as mentioned in Section \ref{sec:sample}.
For the SDSS-DR7 sources, we chose to use the 2500\,\AA\ flux densities provided by \cite{Shen2011}, inferred from a complete spectral fitting procedure that took into account continuum and emission lines for quasar spectra. A detailed description of the analysis of the UV/optical/NIR SED is given in Appendix \ref{sec:appendix1}. Table~\ref{tab2} lists the final 2500-\AA\ fluxes used for the statistical analysis.

In the following, we focus on the analysis in the X-ray band.

\begin{figure}
	\includegraphics[width=1.0\columnwidth]{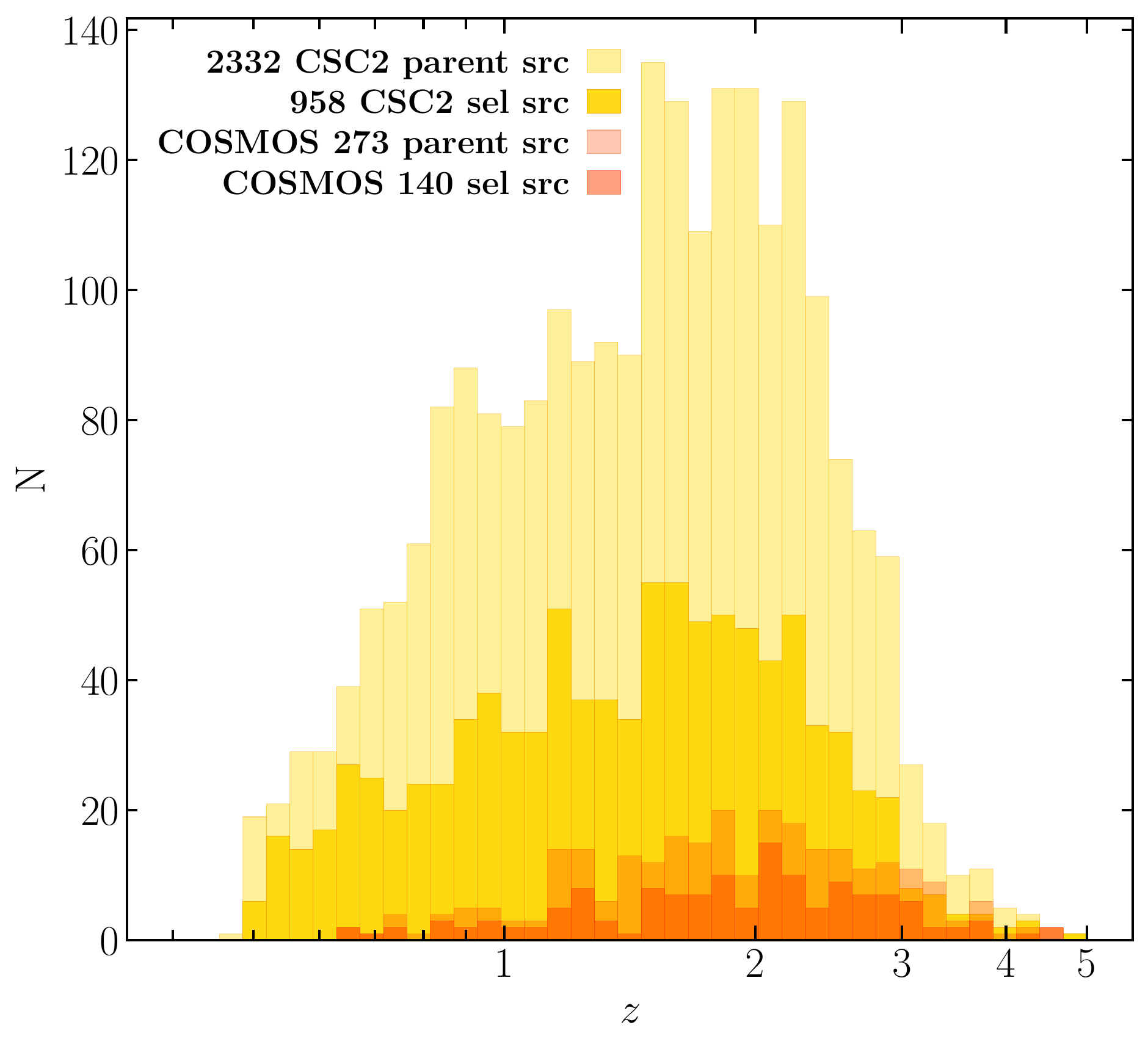}
\caption{Distribution in redshift of the sample of 2,332 SDSS-CSC2.0 and 273 \emph{Chandra}-COSMOS Legacy non-BAL, radio-quiet, dust-free selected sources. The same is shown for the final sample, selected based on the analysis in the X-ray band.}
    \label{fig:dist_z}
\end{figure}

\begin{figure}
	\includegraphics[width=1.0\columnwidth]{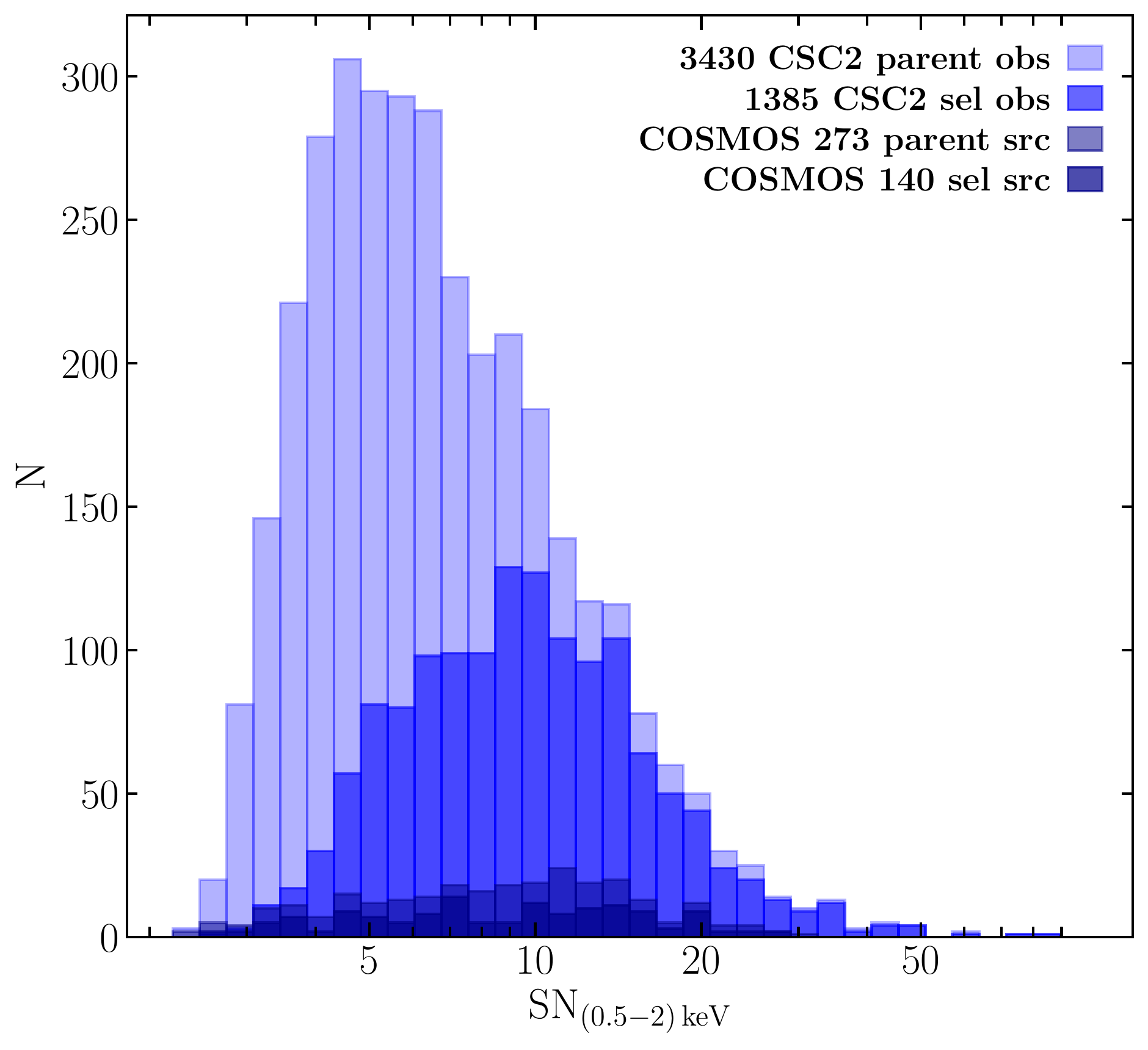}
\caption{Distribution of the signal-to-noise in the soft band for the 3,430 CSC 2.0 observations, corresponding to 2,332 non-BAL, radio-quiet, dust free sources and for the 273 \emph{Chandra} COSMOS Legacy sources. The same is shown for the final sample, selected based on the analysis in the X-ray band.}
    \label{fig:dist_SN}
\end{figure}

\subsection{SDSS-CSC2.0 X-ray spectral analysis} \label{subsec:X_SED}

\subsubsection{Spectral fitting}

\begin{figure}
	\includegraphics[width=1.0\columnwidth]{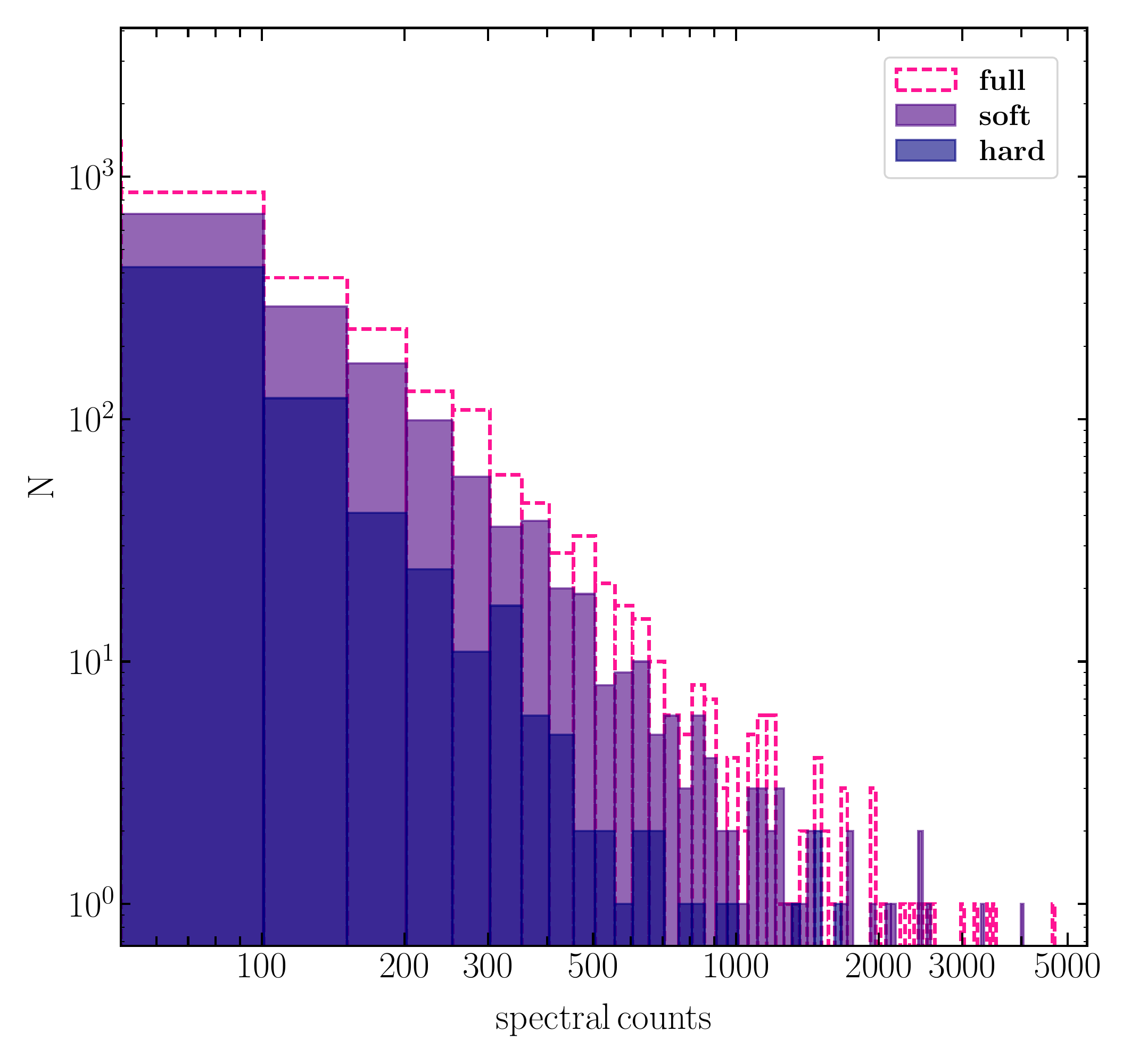}
\caption{Distribution of spectral counts in the full (0.5--7 keV), soft (0.5--2 keV) and hard band (2--7 keV).}
    \label{fig:dist_counts}
\end{figure}

The \emph{CSCview} application provides access to the CSC 2.0 catalogue allowing the user to download uniformly extracted spectra\footnote{Software and calibration database versions used for each observation are specified in the header of the .fits files.} and ancillary files (response matrices and background spectra) for each source in each observation.
A complete spectral analysis was carried out with the fitting package XSPEC v12.10.1b \citep{Arnaud1996} for each source in the selected observations. 
All of the spectra analysed have at least five counts in the soft band, while only 20 have no counts in the hard band. 3,420/3,430 have $>$10 counts in the full band  and $\sim$34\% have $>$100 counts.  Fig. \ref{fig:dist_counts} shows the distributions of the full, soft and hard spectral counts for the sample. No minimum number of counts or of net vs background counts (SN) has been set for performing the fit.

\begin{figure}
	\includegraphics[width=1.0\columnwidth]{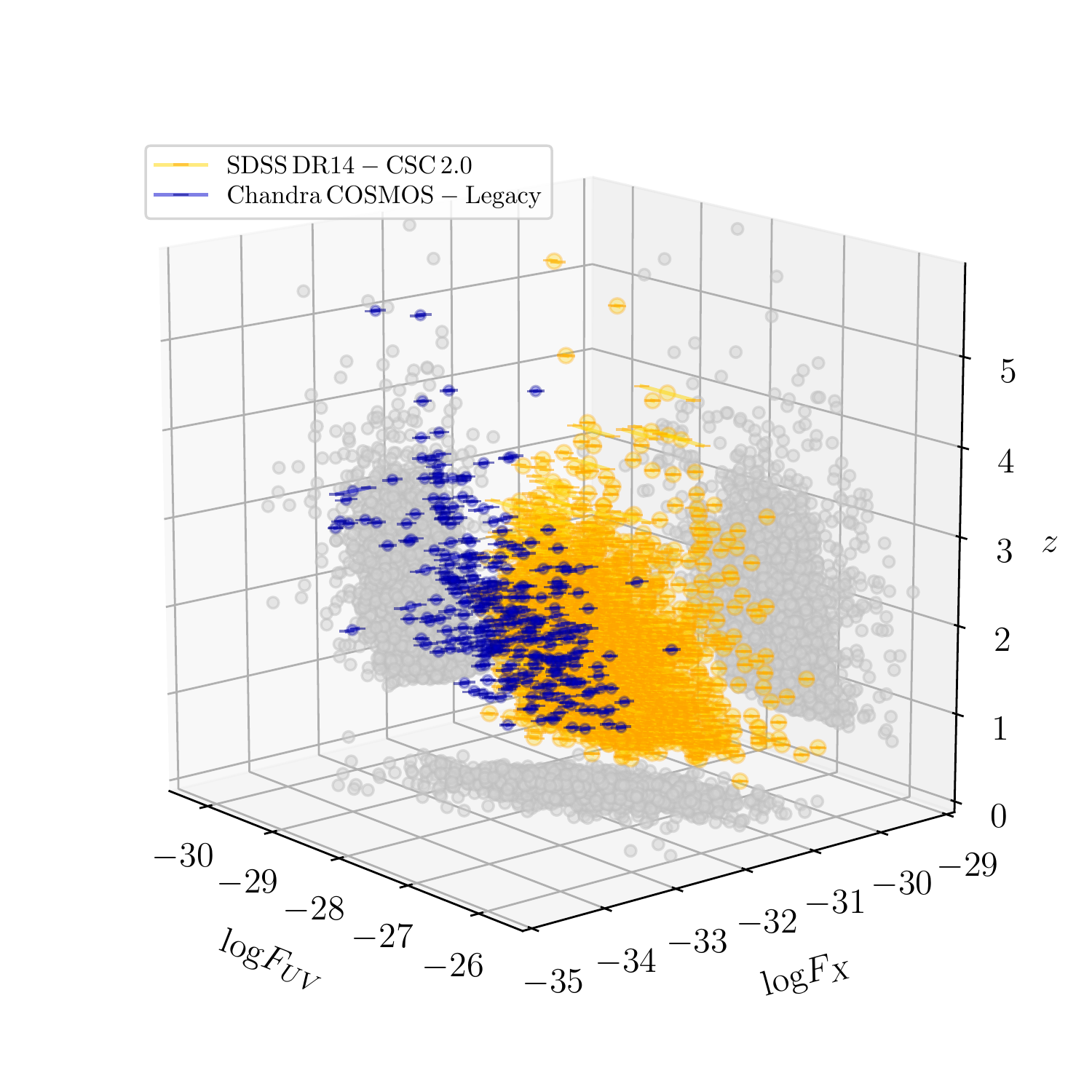}
\caption{Distribution of the 2-keV and 2500-\AA\ fluxes for the 2,332 CSC 2.0 (yellow) and 273 \emph{Chandra} COSMOS Legacy (blue) non-BAL, radio-quiet, dust-free sources. For CSC 2.0 sources with multiple observations, the mean of the log $F_{\rm X}$ values is plotted.}
    \label{fig:flux_dist_z}
\end{figure}

\begin{figure}
	\includegraphics[width=1.0\columnwidth]{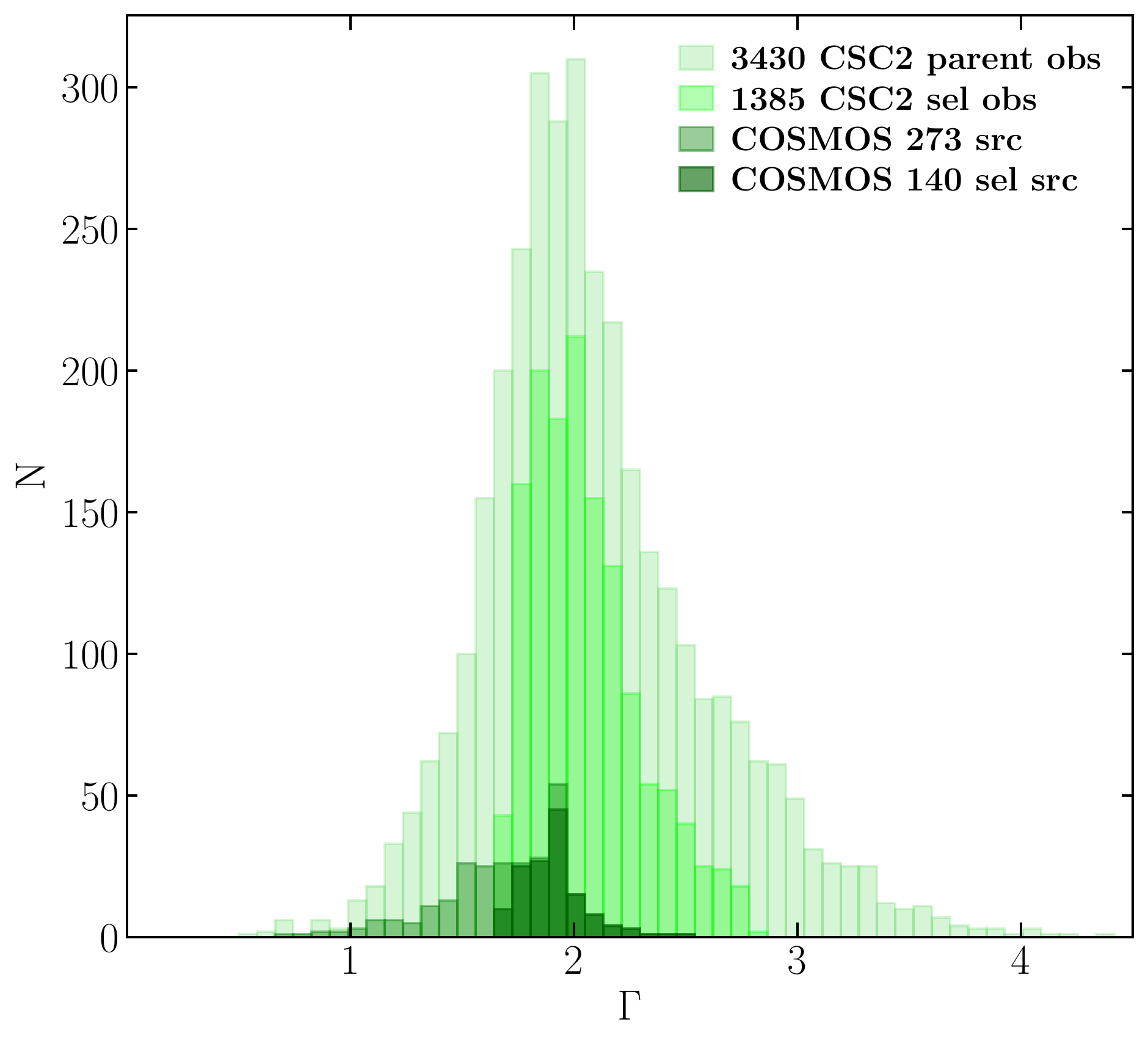}
\caption{Distribution of the photon index $\Gamma$ for the 3,430 selected SDSS-CSC2.0 observations, corresponding to 2,322 non-BAL, radio-quiet, dust-free sources and for the 273 \emph{Chandra} COSMOS Legacy sources. The same is shown after for the final sample, selected based on the analysis in the X-ray band.}
    \label{fig:dist_gamma}
\end{figure}

We assumed a power-law with Galactic absorption $N_{\rm H}$, whose values were drawn from the CSC 2.0 catalog, as a fitting model to the spectrum. 
The best fit to the data allowed the determination of the integrated flux in the soft and hard energy bands ($F_{0.5\mhyphen2 \, \mathrm{keV}}$ and $F_{2\mhyphen7 \,\mathrm{keV}}$), and of the rest-frame flux at 2 keV, tracer of the corona emission, and its error - once the \emph{cflux} component\footnote{The \emph{cflux} (\emph{calculate flux}) model component in XSPEC allowed us to treat the intrinsic power-law flux in a given energy range of interest as a fit parameter (in lieu of the power-law normalisation).} was included in the model - leaving flux and photon index $\Gamma$ as free parameters of the fit.
The flux (luminosity) values at 2 keV were determined with uncertainties of $0.10^{+0.10}_{-0.05}$ dex (median,16$^{\mathrm{th}}$ and 84$^{\mathrm{th}}$ percentile of the error distribution).
Fig. \ref{fig:flux_dist_z} and Fig. \ref{fig:dist_gamma} show the distribution of fluxes at $2$ keV and 2500\,\AA\ as a function of redshift and that of the photon indexes $\Gamma$ for the single observations, respectively, as obtained from the data analysis. 

Due to the presence of sources with an unexpectedly high ($\Gamma>3$) or low ($\Gamma<1.4$) photon index, we checked our spectral analysis looking at the hardness ratios (HR, see Appendix \ref{sec:appendix1bis}). The comparison between the
HRs computed using the net counts from the CSC 2.0 catalog,
and those derived from our
spectral analysis shows an excellent agreement, especially below $\Gamma=2.8$,
the upper threshold we adopted for the photon index.
This is a model-independent confirmation of the reliability of our spectral analysis (see Appendix B for details).

\begin{table*}[ht!]
\scriptsize
\caption{Summary of optical and X-ray properties.}
\begin{center}
\renewcommand{\arraystretch}{1.0}
\begin{tabular*}{1.0\linewidth}{@{\extracolsep{\fill}}l c c c c c c c c c c c c c}
\hline
SDSS name & redshift  & OBSID$^{a}$ &  $\Theta^{b}$ & Exp$^{c}$ &  DRflag$^{d}$  &   $f_{2500}^{e}$& $f_{2\mathrm{keV}}^{f}$&$\Gamma^{g}$&$f_{\mathrm{lim}_{2\mathrm{keV}}}^{h}$& $C_{soft}^{i}$& $C_{hard}^{j}$ & $\mathrm{SN}_{\mathrm{soft}}^{k}$ &$\mathrm{SN}_{\mathrm{hard}}^{l}$  \\
    &   &   &   &   &   &    &   &   &  & & & &     \\
\hline
000102.74+023503.2 &$0.76489$&4837& $4.80644 $ &  $5788.17$ &   DR14  &  $-28.21\pm0.02$   & $-31.12_{-0.07}^{+0.06}$     &   $1.7_{-0.2}^{+0.3}$  & $-31.79$  & $42$ & $22$  &  $6.4$ & $4.6$  \\
000130.63+233443.5&$2.93$       &14898 & $5.82812$ & $39465.1$&    DR12     & $-28.01\pm0.05$    & $-32.04_{-0.12}^{+0.11}$     &$1.5_{-0.2}^{+0.2}$     &  $-32.80$  & $63$ & $34$ & $7.4$ & $4.9$     \\
 000150.39+023830.2&$0.949$    & 4837 &  $8.79053$ &  $5788.17$&    DR12  &  $-27.833\pm0.010$   & $-31.20_{-0.09}^{+0.08}$     & $3.4_{-0.5}^{+0.5}$    &  $-31.59$ & $28$ & $8$  &  $5.2$  & $2.4$  \\
000151.58+232857.6&$0.884201$&14898 & $2.47162$ &  $39461.9$&    DR14  & $-27.747\pm0.014$    & $-31.74_{-0.06}^{+0.06}$     &$2.0_{-0.2}^{+0.2}$     &   $-32.48$  & $60$ & $25$  & $7.7$ & $4.9$   \\
 000151.95+232421.0&$2.03318$&14898  & $5.64085$ & $39461.9$  &      DR12   &  $-28.46\pm0.05$   &  $-31.99_{-0.13}^{+0.13}$    &  $2.2_{-0.3}^{+0.4}$   &  $-32.52$ &$30$ & $12$    & $5.2$ &$2.5$   \\

\hline
\end{tabular*}
     \vspace{1ex}

     {\raggedright {\bfseries Notes.} Summary of optical and X-ray properties inferred from the analysis for the first five observations in the sample. 
     This table is available in its entirety at the CDS. 
     
     $^{a}$\emph{Chandra} observation identifier. $^{b}$Off-axis angle [arcmin] $^{c}$Exposure time [s]. $^{d}$SDSS Data Release. $^{e}$Flux at $2500$\AA\ as inferred from the UV SED analysis [erg s$^{-1}$cm$^{-2}$Hz$^{-1}$]. $^{f}$Flux at 2 keV as inferred from the spectral analysis assuming a power-law model and Galactic absorption $N_{\rm H}$ [erg s$^{-1}$cm$^{-2}$Hz$^{-1}$]. $^{g}$Photon index. $^{h}$Flux limit at 2 keV per observation [erg s$^{-1}$cm$^{-2}$Hz$^{-1}$]. $^{i}$Raw counts in the soft band (0.5--2 keV). $^{j}$Raw counts in the hard band (2--7 keV). $^{k}$Signal-to-noise in the soft band (0.5--2 keV). $^{l}$Signal-to-noise in the hard band (2--7 keV). \par}

\label{tab2}
\end{center}
\end{table*}

 \subsubsection{Computation of the flux limit}
Active galactic nuclei with an X-ray flux measurement close to the flux limit of the observation will be observed only in case of a positive fluctuation. This well known effect, named Eddington bias, introduces a systematic and also redshift dependent bias at the faint end of the X-ray flux distribution, which has the effect  to flatten the $\Lx-\Lo$ relation (see also Section 5.3 of \citealt{lusso2020}). We therefore computed the rest-frame 2 keV flux limit for each observation to exclude the sources caught just as a positive fluctuation. This flux was computed by interpolating - or extrapolating, depending on the redshift of the source - the monochromatic flux limits $f_{\mathrm{S}}$ and $f_{\mathrm{H}}$ in the broad soft and hard bands. 
This same method was used for inferring the rest-frame $2$ keV flux when only photometric data are available, and was employed in the case of the \emph{Chandra} COSMOS Legacy sample (see Section \ref{sec:COSMOSphot} for a detailed explanation of the method).

The flux limit was then derived as follows: for each observation, we made use of the percentage of net counts - measured by XSPEC as the number of source over total (source+background) counts - in the soft $0.5-2$ keV  and hard $2-7$ keV bands to compute the significance of the source ($S_{soft}$ and $S_{hard}$), and a factor taking into account the background level in the integrated bands, $P_{bkg} = \frac{1-S}{S}$. 
By multiplying $F_{0.5\mhyphen2 \,\mathrm{keV}}$ and $F_{2\mhyphen7 \,\mathrm{keV}}$ for $P_{bkg}$, we obtained an approximation of the background flux in the soft and hard band respectively. In doing so, even if we were incorrectly assuming the same spectral shape for source and background, the background flux in the soft/hard band turned out to be slightly higher than the one obtained with the less steep spectral shape characterising the background. Since, as explained below, we were indirectly using this value to exclude from the sample the sources close to the flux limit - proportional to the background flux - a higher value resulted in a more conservative selection.
The monochromatic background fluxes $f_{bkg \,\mathrm{S}}$ and $f_{bkg \,\mathrm{H}}$ in the soft and hard bands were then used to infer the flux limit in the bands, as shown by the following expression for the soft-band case:
\begin{equation} \label{eq:flim}
f_{lim_{\mathrm{S}}} = \frac{\eta_{\mathrm{S}} \, \mathrm{SN}_{\mathrm{MIN}}^{2} + \mathrm{SN}_{\mathrm{MIN}} \, \sqrt{\eta_{\mathrm{S}}^{2}\,\mathrm{SN}_{\mathrm{MIN}}^{2} + 8\, \eta_{\mathrm{S}} \, f_{bkg_{\mathrm{S}}}}}{2} \,,
\end{equation}
where $\eta_{\mathrm{S}}$, the conversion factor between counts and fluxes ($\eta \,[\mathrm{erg s}^{-1} \mathrm{cm}^{-2} \mathrm{Hz}^{-1} \mathrm{counts}^{-1}] = f / C$), is given by
\begin{equation} \label{eq:kfactor}
\eta_{\mathrm{S}} = \frac{f_{\mathrm{S}} + \sqrt{f_{\mathrm{S}}^{2}+8\,\mathrm{C}_{\mathrm{raw \, \mathrm{S}}} \, \Delta f_{\mathrm{S}}^{2}}}{4\,\mathrm{C}_{\mathrm{raw \, \mathrm{S}}}} \,,
\end{equation}
where $\mathrm{C}_{\mathrm{raw \, \mathrm{S}}}$ includes net and background counts,
and both equations (\ref{eq:flim}) and (\ref{eq:kfactor}) are obtained, after simple algebra, from the equation
\begin{equation}
\mathrm{SN}_{\mathrm{MIN}} = \frac{\mathrm{C}_{\mathrm{MIN}}}{\sqrt{\mathrm{C}_{\mathrm{MIN}}+2\,\mathrm{C}_{bkg}}}\,,
\end{equation}
where the choice of a minimum signal-to-noise in the formula, $\mathrm{SN}_{\mathrm{MIN}}$ (in our case $\mathrm{SN}_{\mathrm{MIN}}=3$), implies the other quantities. 
The $f_{lim \, 2\mathrm{keV}}$ was then inferred from interpolation or extrapolation of $f_{lim \, \mathrm{S}}$ and $f_{lim \, \mathrm{H}}$, depending on the redshift of the source.

The X-ray properties inferred from this analysis are listed in Tab.~\ref{tab2}.

\subsection{COSMOS X-ray photometric analysis}\label{sec:COSMOSphot}

We decided to consider the background-subtracted, aperture-corrected fluxes values already listed in the \emph{Chandra} COSMOS Legacy catalogue. The photometric data were estimated from the count rates $R$ in each band -  obtained with the \emph{Chandra} \texttt{Emldetect} (\texttt{CMLDetect}) maximum likelihood algorithm - using the relation $F = R \times (CF \times 10^{-11})$, with $CF$ energy conversion factor. $CF$ was computed with the CIAO tool \emph{srcflux}, assuming a power-law spectrum with $\Gamma$ = 1.4 and a Galactic column density of $\mathrm{N}_{\mathrm{H}} = 2.6 \times 10^{20}$ cm$^{-2}$ \citep[for details, see][]{Civano2016}.
 
To infer the flux at rest-frame 2 keV, we interpolated - or extrapolated, depending on the redshift of the source - the monochromatic flux in the broad soft and hard bands that we estimated from the integrated values available from the \emph{Chandra} COSMOS Legacy catalogue. 
 In order to choose the energies in the soft and hard bands at which computing the monochromatic fluxes, we applied a new method developed by our group \citep[][see Section 1 of their Supplementary Material]{RisalitiLusso2019}. Instead of using the more conventional geometric mean or the mean energy of a band, we defined the pivot point $E_{p}$ as the specific energy 
at which the monochromatic flux $f(E_{p})$ and the photon index $\Gamma$ of the spectrum have a null covariance, that is the errors on the two quantities are independent. To compute this energy, a spectrum of high signal-to-noise was simulated, assuming typical calibration and an average background for \emph{Chandra} observations, along with a typical photon index of $1.7$ for the spectrum. We performed fits to the simulated data for different values of the energy $E_{p}$, assuming a spectrum $f(E)=f(E_{p})(E/E_{p})^{\Gamma}$ and $f(E_{p})$ and $\Gamma$ as free parameters.
Error contours of $f(E_{p})$ versus $\Gamma$ were plotted within XSPEC. When the two quantities are co-dependent, the ellipsoid of the error contours appears tilted with respect to the $x$- and $y$-axis, as expected when the covariance is different from zero. When, instead, the axes of the ellipsoid are parallel to the $x$ and $y$ axes, the errors are independent. We found the values at which this occurs to be $E_{\mathrm{S}}=1.05$ keV and $E_{\mathrm{H}}=3.1$ keV for the soft and the hard band respectively.
The pivot energies have many advantages: 1) by definition, at these energies the covariance between  $f(E_{p})$ and $\Gamma$ is zero; 2) the flux $f(E_{p})$ is independent from the value of $\Gamma$ assumed for the energy spectrum in the band, that is the accuracy of the flux estimate is independent from the value assumed by the photometric catalogue used (for CSC 2.0, $\Gamma=2$); 3) following the relation between the total flux in the band ($F$) and at a monochromatic energy ($f(E)$)\footnote{$F=\int_{E_{1}}^{E_{2}} f(E) dE$}, we can write:
\begin{equation}
f(E) = \frac{(2-\Gamma) E^{1-\Gamma}}{E_{2}^{2-\Gamma}-E_{1}^{2-\Gamma}} F\,.
\end{equation}
When $E=E_{p}$, the relative error on the monochromatic flux is the same as the one in the band , that is $\Delta(f(E_{p}))/f(E_{p}) = \Delta(F)/F$, and the absolute error $\Delta(f(E_{p}))$ is the smallest possible. 
    We then computed the value of the rest 2-keV flux by interpolation (or extrapolation) of the monochromatic soft and hard fluxes estimated at the pivot frequencies from the integrated fluxes in the bands. When only the soft band was available, we inferred the 2-keV flux by assuming a typical photon index $\Gamma=1.9$. A comparison of the COSMOS and SDSS-CSC2.0 rest-frame 2 keV and 2500\,\AA\ fluxes as a function of redshift is shown in Fig. \ref{fig:flux_dist_z}. The \emph{Chandra} COSMOS Legacy sample (blue data points) covers a fainter locus of the distribution between the fluxes ($\langle\mathrm{log} f_{\rm 2 keV}\rangle = -32.40$, $\langle\mathrm{log} f_{\rm 2500 \AA}\rangle = -28.92$ versus $\langle\mathrm{log} f_{\rm 2 keV}\rangle = -31.59$, $\langle\mathrm{log} f_{\rm 2500 \AA}\rangle = -27.82$ of the SDSS-CSC2.0 sample), and, on average, a slightly higher redshift ($\langle z \rangle \sim 1.8$ vs. $\langle z \rangle \sim 1.5$).
The same procedure was employed for the determination of the COSMOS flux limit at 2 keV rest-frame. This time, flux limits in the soft and hard bands were already available from the sensitivity maps of the survey.

\subsection{Eddington bias threshold estimation}
\begin{figure}
	\includegraphics[width=1.0\columnwidth]{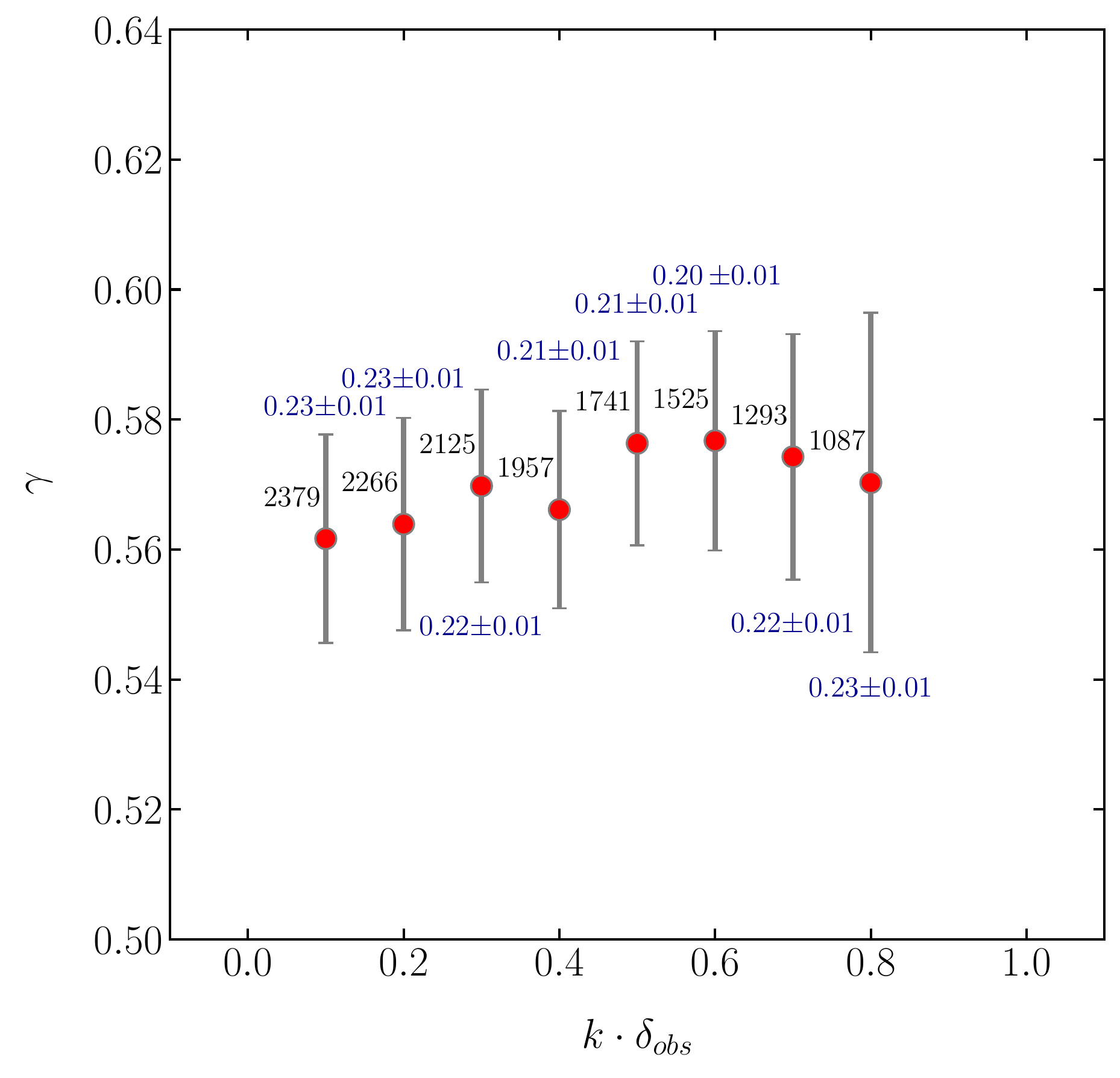}
\caption{Mean slope for the $f_{\rm X}-f_{\rm UV}$ relation for the case of $\Gamma=1.7$--2.8, evaluated as the arithmetic mean of all the slopes in the redshift bins with more than five objects, as a function of $k  \, \delta_{obs}$, that is of the amount of sources excluded from the sample because of the selection for the Eddington bias (see text for details). The mean slope is flatter for low values of $k$, reaches a maximum of $\gamma \sim 0.58$ and then drops back while $k$ increases. 
Mean dispersion as a function of $k  \, \delta_{obs}$ is listed in blue, while in black the number of sources surviving the selection for each $k  \, \delta_{obs}$ choice.}
    \label{fig:edd_bias}
\end{figure}

A quasar whose average X-ray emission is close to the flux limit of the observation will be detected only if caught on a positive fluctuation (Eddington bias). The inclusion of such sources corresponds to the presence of non-representative data points in the $L_{\rm X}-L_{\rm UV}$ relation.
In order to clean the sample from this effect, we excluded those sources with an \emph{expected} X-ray flux lower than a threshold defined by the flux limit of their observation. This was done as follows:
1.) from the $L_{2500}$ obtained through the UV SED analysis and the assumption of the concordance $\Lambda$CDM model, we estimated the expected $L_{\rm X}$, assuming slope and intercept of the $L_{\rm X}-L_{\rm UV}$ relation from \cite{LussoRisaliti2016};
2.) we derived the expected X-ray flux and define a threshold $f_{thr}$. We kept in the sample only those sources for which
\begin{equation}
\mathrm{log}(f_{lim}) < \mathrm{log}(f_{thr}) - k \,\delta_{obs}\,,
\end{equation}
where $k$ is a number and $\delta_{obs}$ is the observed dispersion in the sample.
The number of rejected sources depends on the choice of $k$, which we evaluated with the following procedure:
we divided the sample in redshift bins (see details in Section~\ref{sec:analysis}) and analysed the relation between the fluxes in the UV and X-ray bands. If the size of the redshift bin is small enough to make the difference in the luminosity distances negligible, that is if the contribution to the dispersion in the relation due to the difference in the luminosity distance among the objects is smaller than the observed $\delta$, we can use the flux as a proxy for the luminosity and examine the $f_{\rm X}-f_{\rm UV}$ relation, making our analysis cosmology-independent.
The inclusion of sources in the sample having a positive fluctuation of their X-ray emission above the flux limit, but with an observed X-ray flux value very close to the flux limit of the given observation, causes a `flattening' of the slope in the relation between the fluxes. 
We therefore analysed the behaviour of the mean slope and dispersion for the $f_{\rm X}-f_{\rm UV}$ relation - in all the redshift bins with a number of sources $>5$ - at increasing values of $k$, that is at more conservative cuts for the Eddington bias. This was done as follows: 1. we divided the sample in redshift bins with size $\Delta\mathrm{log}\, z =0.06$ (similar results are found for $\mathrm{log}\, z =0.05$ and $\mathrm{log}\, z =0.07$); 2. we computed the fit to the data for the $f_{\rm X}-f_{\rm UV}$ relation, obtaining a slope and intercept, in each redshift bin; 3. we then computed the arithmetic mean for all the slopes in the bins, obtaining an average value. We then repeated the steps above for different choices of $k \, \delta_{obs}$, with the intention of determining the $k \, \delta_{obs}$ value for which the trade-off between number of sources kept in the sample and the flattening of the relation between $\gamma$ and $k \, \delta_{obs}$ was met. Fig. \ref{fig:edd_bias} show the eight average slopes result of this analysis for a range of  $k \, \delta_{obs}$ values from $0.1$ to $0.8$.
This analysis was also performed for different choices of the photon index. Here we show the results for the final choice of the range of accepted photon index values ($\Gamma=1.7$--2.8, see Appendix~ \ref{sec:appendix2} for details). 

As shown in Fig. \ref{fig:edd_bias} for the case of $\Gamma=1.7$--2.8, the mean slope of the relation is flatter for low values of $k$, reaches a maximum at $\gamma \sim 0.57$--0.58 and then drops back while $k$ increases. The statistical significance of the difference among the slopes based on the $k \, \delta_{obs}$ value is small, given their uncertainties, meaning that the presence of the Eddington bias in \emph{Chandra} observations is softened by a very low, basically zero, level of background. This represents a marked difference with respect to the SDSS-XMM sample, for which the dependence of the mean slope of the $f_{\rm X}-f_{\rm UV}$ relation in the redshift bins  on the chosen cut in $k  \, \delta_{obs}$ was much stronger \citep{RisalitiLusso2019}.
We chose $k  \, \delta_{obs} = 0.6$ as the best compromise between a steady value for the slope of the relation and the number of sources included in the sample in all the cases we examined.

\section{Statistical analysis: the $\Lx-\Lo$ relation} \label{sec:analysis}

We performed a complete analysis to understand which combination of selection filters in photon index, that is lack of obscuration in the X-rays, and flux threshold, namely Eddington bias, represents the best compromise between the size of the sample and the minimisation of the observed dispersion (see discussion in Appendix~\ref{sec:appendix2}).
Our final choice ($\Gamma=1.7$--2.8, $k  \, \delta_{obs} = 0.6$) reduced the sample to 1,385 CSC 2.0 observations (i.e. less than a half of the original 3,430 observations) and 140 Chandra COSMOS Legacy sources, corresponding to a total of 1,098 single sources (see Tab. \ref{tab1}). When a source was observed more than once, we chose to compute the X-ray flux as a simple arithmetic mean of the multiple measures (see discussion below).

\begin{table*}[ht!]
\footnotesize
\caption{Results of the fitting to the final selected sample.}
\begin{center}
\renewcommand{\arraystretch}{1.0}
\begin{tabular*}{1.0\linewidth}{@{\extracolsep{\fill}} c c c c c c c c}
\hline
$\langle z \rangle$ &  N   &  $\gamma_{\it{emcee}}$   &  $\beta_{\it{emcee}}$   &  $\delta_{\it{emcee}}$      &  $\gamma_{\it{linmix}}$  &  $\beta_{\it{linmix}}$  &    $\delta_{obs}$ \\
           &       &                                         &                                     &                                          &                                   \\
\hline
$0.52$ &   $24$    &  $0.73 \pm 0.15$  &  $-31.3 \pm 0.07$   &  $0.27 \pm 0.05 $ &    $0.73_{-0.15}^{+0.15}$   & $-31.3_{-0.08}^{+0.08}$  &  $  0.26 \pm 0.04$ \\
$0.60$ &   $34$    &  $0.65 \pm 0.07$  &  $-31.3 \pm 0.04$   &  $0.22 \pm 0.03 $ &    $0.65_{-0.08}^{+0.08}$   & $-31.3_{-0.04}^{+0.04}$  &  $  0.22 \pm 0.03$  \\
$0.68$ &   $56$    &  $0.60 \pm 0.07$  &  $-31.3 \pm 0.03$   &  $0.23 \pm 0.03 $ &    $0.60_{-0.07}^{+0.07}$   & $-31.3_{-0.04}^{+0.04}$   &  $  0.23 \pm 0.02$   \\
$0.78$ &   $48$    &  $0.69 \pm 0.08$  &  $-31.3 \pm 0.04$   &  $0.26 \pm 0.03 $ &    $0.69_{-0.08}^{+0.08}$   & $-31.3_{-0.04}^{+0.04}$  &  $  0.26 \pm 0.03$     \\
$0.89$ &   $67$    &  $0.56 \pm 0.08$  &  $-31.4 \pm 0.03$   &  $0.24 \pm 0.02 $ &    $0.56_{-0.08}^{+0.08}$   & $-31.4_{-0.03}^{+0.03}$  &  $  0.25 \pm 0.02$     \\
$1.02$ &   $77$    &  $0.53 \pm 0.07$  &  $-31.4 \pm 0.03$   &  $0.25 \pm 0.02 $ &    $0.53_{-0.07}^{+0.07}$   & $-31.4_{-0.03}^{+0.03}$  &  $  0.25 \pm 0.02$    \\
$1.16$ &   $98$    &  $0.55 \pm 0.06$  &  $-31.5 \pm 0.03$   &  $0.26 \pm 0.02 $ &    $0.55_{-0.06}^{+0.06}$   & $-31.5_{-0.03}^{+0.03}$  &  $  0.27 \pm 0.02$    \\
$1.33$ &   $84$    &  $0.56 \pm 0.06$  &  $-31.5 \pm 0.03$   &  $0.26 \pm 0.02 $ &    $0.56_{-0.06}^{+0.06}$   & $-31.5_{-0.03}^{+0.03}$   &  $  0.27 \pm 0.02$     \\
$1.53$ &  $117$    &  $0.56 \pm 0.05$  &  $-31.6 \pm 0.02$   &  $0.25 \pm 0.02 $ &    $0.56_{-0.05}^{+0.05}$   & $-31.6_{-0.03}^{+0.03}$  &  $  0.26 \pm 0.02$   \\
$1.74$ &  $117$    &  $0.52 \pm 0.04$  &  $-31.6 \pm 0.02$   &  $0.21 \pm 0.02 $ &    $0.52_{-0.05}^{+0.05}$   & $-31.6_{-0.02}^{+0.02}$  &  $  0.23 \pm 0.02$     \\
$1.99$ &  $123$    &  $0.48 \pm 0.04$  &  $-31.7 \pm 0.02$   &  $0.22 \pm 0.02 $ &    $0.48_{-0.04}^{+0.04}$   & $-31.7_{-0.02}^{+0.02}$  &  $  0.24 \pm 0.02$     \\
$2.28$ &   $104$    &  $0.55 \pm 0.03$  &  $-31.7 \pm 0.02$   &  $0.19 \pm 0.02 $ &    $0.55_{-0.03}^{+0.03}$   & $-31.7_{-0.02}^{+0.02}$   &  $  0.22 \pm 0.02$     \\
$2.6$  &   $74$    &  $0.56 \pm 0.05$  &  $-31.8 \pm 0.03$   &  $0.23 \pm 0.02 $ &    $0.56_{-0.05}^{+0.05}$   & $-31.8_{-0.03}^{+0.03}$  &  $  0.25 \pm 0.02$    \\
$2.98$ &   $43$    &  $0.47 \pm 0.05$  &  $-31.8 \pm 0.05$   &  $0.24 \pm 0.04 $ &    $0.47_{-0.05}^{+0.05}$   & $-31.8_{-0.05}^{+0.05}$  &  $  0.27 \pm 0.03$     \\
$3.4$  &   $15$    &  $0.64 \pm 0.07$  &  $-31.7 \pm 0.06$   &  $0.17 \pm 0.06 $ &    $0.64_{-0.08}^{+0.08}$   & $-31.7_{-0.08}^{+0.07}$  &  $  0.20 \pm 0.04$     \\
$3.89$ &   $9$     &  $0.58 \pm 0.07$  &  $-31.8 \pm 0.05$   &  $0.00 \pm 0.01 $ &    $0.58_{-0.09}^{+0.10}$   & $-31.8_{-0.07}^{+0.08}$  &  $  0.06 \pm 0.01$     \\
$4.45$ &   $6$     &  $0.58 \pm 0.13$  &  $-31.8 \pm 0.15$   &  $0.15 \pm 0.15 $ &    $0.56_{-0.24}^{+0.24}$   & $-31.8_{-0.23}^{+0.22}$   &  $  0.21 \pm 0.06$     \\
\hline
\end{tabular*}
     {\raggedright {\bfseries Notes.} Results of the fitting to the final selected sample ($\Gamma=1.7$--2.8, $k  \, \delta_{obs}=0.6$) in the 17 redshift bins of size $\Delta \mathrm{log} (z) = 0.06$. The fitting was performed with the two Bayesian MCMC methods \emph{emcee} and \emph{linmix} (see the text for details). \par}
\label{tab3}
\end{center}
\end{table*}

The statistics and the quality of the data involved allowed an examination of the behaviour of the relation with redshift within narrow, and yet well populated, redshift bins.
This test served two fundamental purposes. On the one hand, checking the non-evolution of the slope in the relation between the fluxes, the one we are relying on for measuring cosmological distances, is crucial for the cosmological employment of the relation. On the other hand, it gives us the opportunity of examining the physics connecting accretion disc and hot corona up to a redshift at which AGN were younger, and see if the process linking the two components appears to be the same over time.

\begin{figure*}
	\includegraphics[width=2.0\columnwidth]{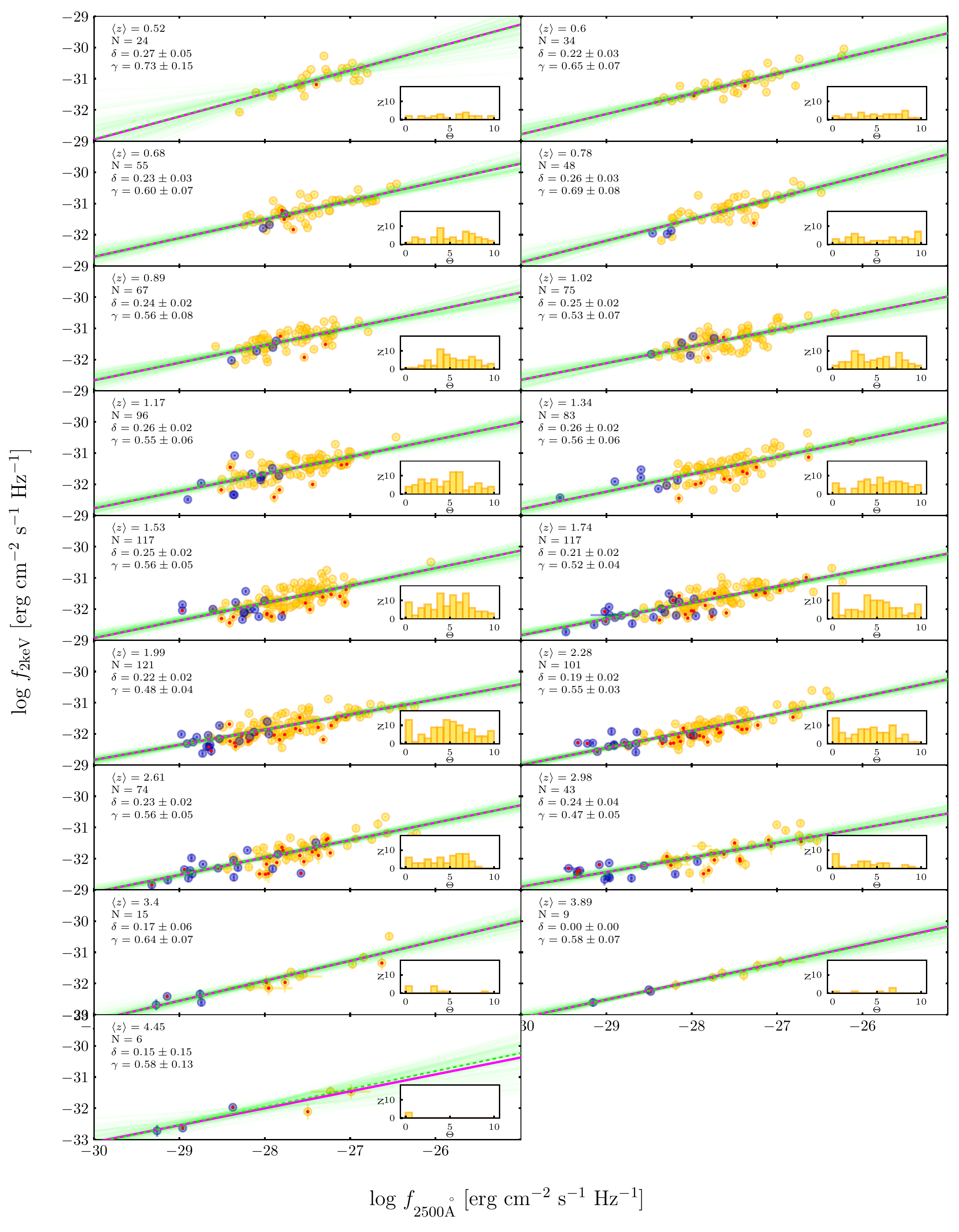}
\caption{Regression analysis of the final selected sample ($\Gamma=1.7$--2.8, $k  \, \delta_{obs}=0.6$). The division of the sample in logarithmic redshift bins of size $\Delta \mathrm{log}(z)=0.06$ yields 17 bins with more than five objects. The results for two fitting methods adopted, \emph{emcee} and \emph{linmix}, are plotted in (dashed) lime green and magenta, respectively. For each sub-plot, the inset shows the distribution of off-axis angles $\Theta$ for the sources in the redshift bin. Sources with at least one observation with SN<5 in the soft band are marked with a red point. For each redshift bin, we list the median redshift $\langle z \rangle$, the number of data points $N$ and dispersion and slope from the \emph{emcee} algorithm.}
    \label{fig:zbins}
\end{figure*}

When the size of the logarithmic redshift bin is small enough, we can use fluxes in place of luminosities, performing a test on the \mbox{(non-)evolution} with redshift that is completely independent from any assumption on cosmology.
\cite{RisalitiLusso2019} analysed in detail the choice of the bin size and verified that, as long as $\mathbf{\Delta}$log$(z)\leq0.1$, the slope in the relation does not depend on it. Thanks to the statistics available, we chose $\Delta$log$(z) =0.06$ and we limited our analysis to the redshift bins with more than five objects. We performed the same analysis for bins of size $\Delta$log$(z) =0.05$ and $\Delta$log$(z) =0.07$, finding no significant difference (see Fig. \ref{fig:recap_zbins} for a comparison of the results for different sizes of the redshift bins).
For our 
selection in $\Gamma$ and the choice of the threshold for the Eddington bias, the division yielded 17 redshift bins.
To perform the fitting to the data, we adopted the Python package \emph{emcee} \citep{Foreman2013}, a pure-Python implementation of Goodman \& Weare's Affine Invariant Markov Chain Monte Carlo (MCMC) Ensemble sampler. To check that the results were independent from the employed method, we also performed the analysis using the \emph{Linmix} package \citep{Kelly2007}, an algorithm that makes use of a Bayesian approach to linear regression and takes into account the errors in both the $x$ and $y$ variable.
We performed the fit a first time, then applied a $3 \sigma$ clipping to the data, repeating this sequence for a total of three times. This yielded no significant difference with respect to the analysis without $\sigma$ clipping.
The results are shown in Fig.~\ref{fig:zbins} and Fig.~\ref{fig:recap_zbins}, and summarised in Tab.~\ref{tab3}. In Fig.~\ref{fig:zbins}, red points indicate when the observations are characterised by a SN<5 in the soft band. Most of them are (X-ray) fainter objects at intermediate redshifts. 
This confirms that data points that passed our selection criteria, even if with a low SN, follow the relation and are representative of the population of blue quasars.

\begin{figure}
\begin{center}
	\includegraphics[width=1.\columnwidth]{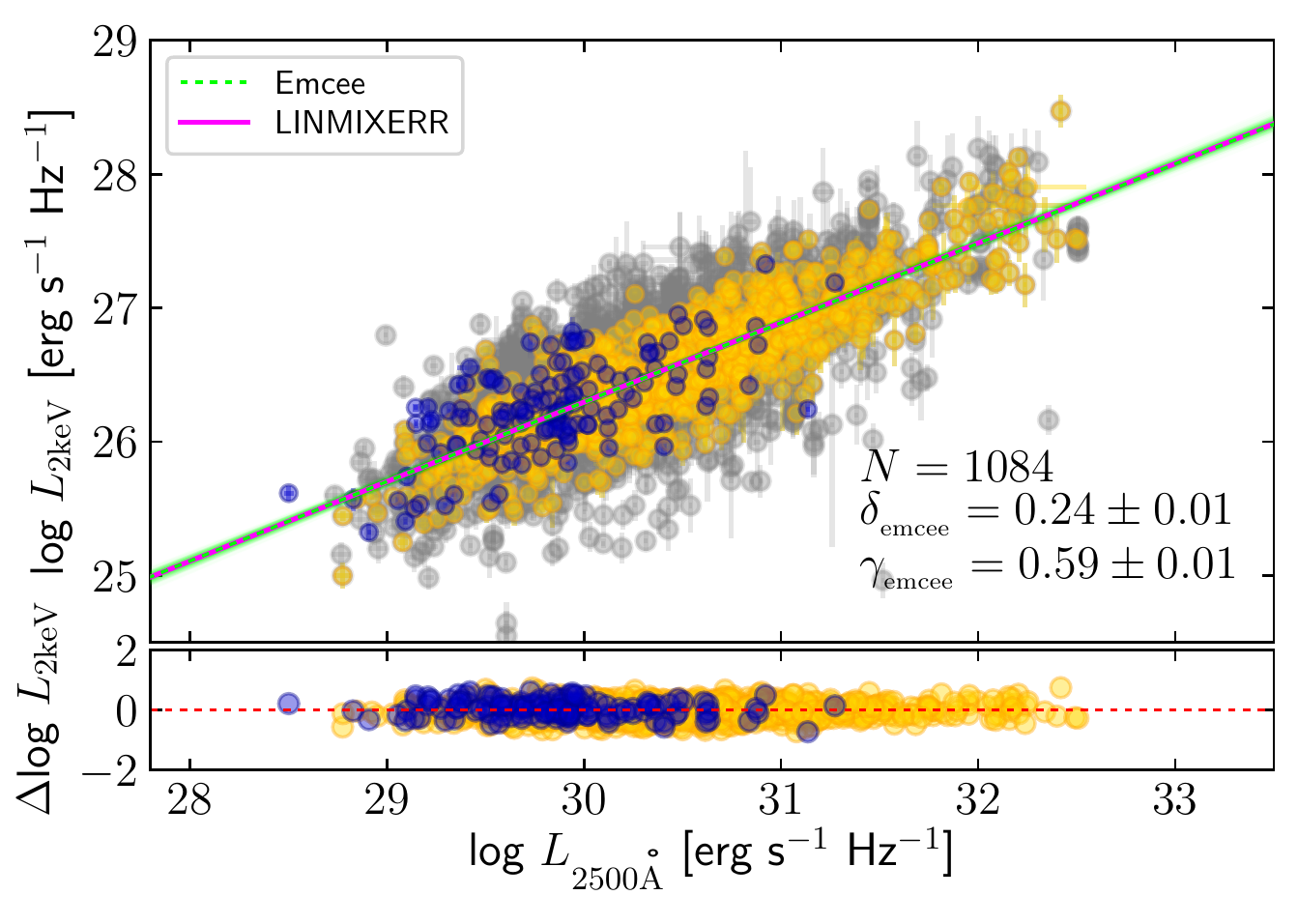}
\caption{Regression analysis of the $\Lx-\Lo$ relation for the final, entire selected SDSS-CSC2.0 (yellow) and \emph{Chandra} COSMOS Legacy (blue) sample ($\Gamma=1.7$--2.8, $k  \, \delta_{obs}=0.6$), while in grey is shown the pre-selected parent sample (2,332 SDSS-CSC2.0 (3,430 observations) and 273 \emph{Chandra} COSMOS Legacy sources), before the selection in the X-ray band is applied (the observed dispersion in this case is $\delta_{obs}=0.32$ dex). The use of spectroscopic data leads to a final dispersion of $0.24$ dex, in agreement with the one found in previous works on samples of similar size using photometric data.}
    \label{fig:lo_lx}
    \end{center}
\end{figure}

For the relation between the luminosities on the selected sample, we adopted the same fitting methods and applied a $3 \sigma$ clipping as described above. The result is shown in Fig.~\ref{fig:lo_lx}, where the final selection of CSC 2.0 and COSMOS sources is in gold and blue, respectively, while the original sample of 3,430 observations with a $\delta_{obs}=0.32$ dex, is in grey.
The exclusion of 14/1,098 sources at more than $3 \sigma$ from the best-fit led to a final dispersion $\delta = 0.24\pm0.01$ (compared to $\delta = 0.25\pm0.01$ if no sigma-clipping is applied), the same found by \cite{RisalitiLusso2019} with the SDSS DR7/DR12 - \emph{XMM-Newton} sample. The slope is slightly flatter compared to the one of the \emph{XMM} sample ($\gamma = 0.59 \pm 0.01$ vs $\gamma = 0.633 \pm 0.002$, \cite{RisalitiLusso2019}).

\section{Discussion} \label{sec:discussion}

\begin{figure*}
\begin{center}

\includegraphics[width=2.0\columnwidth]{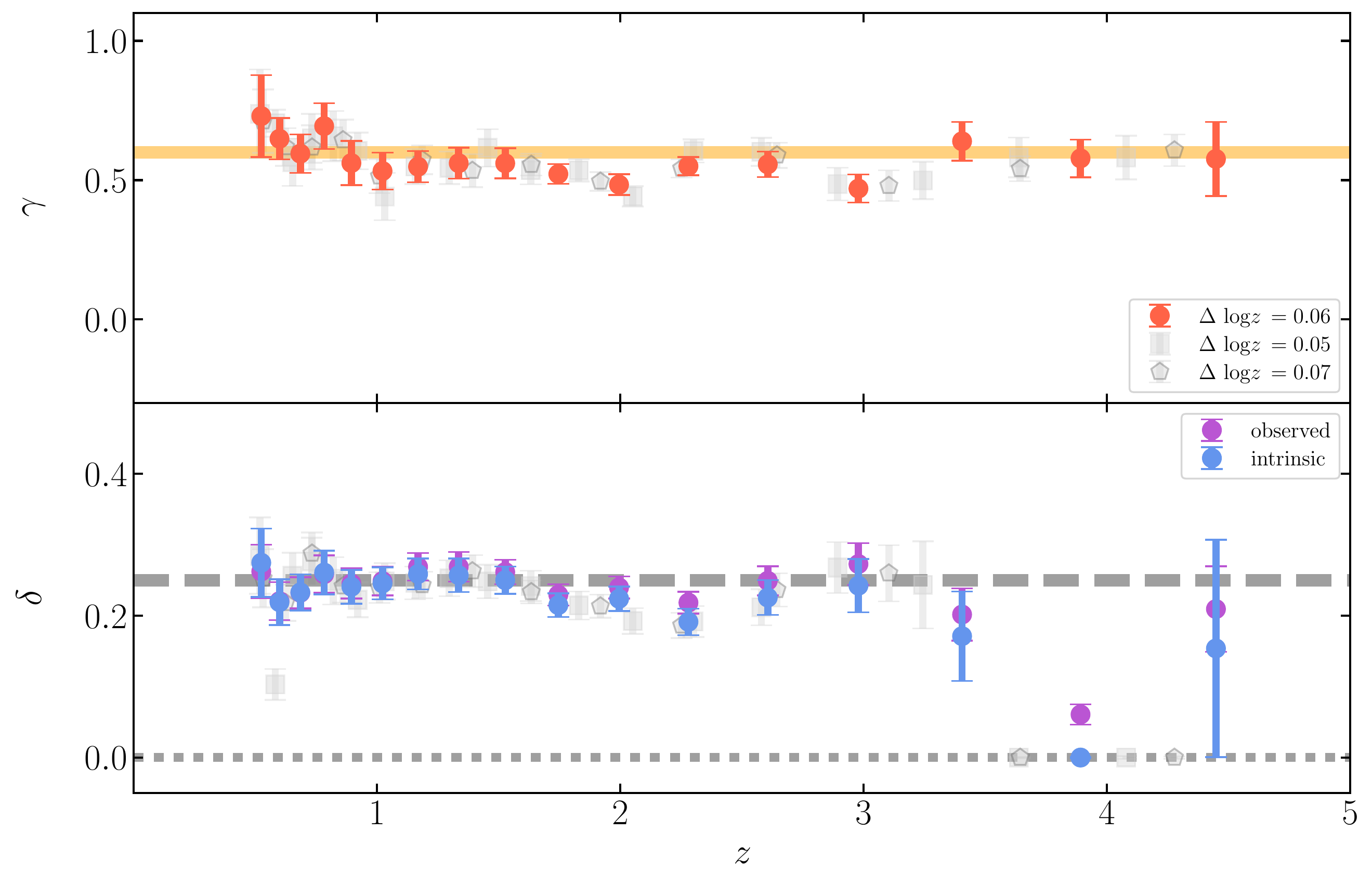}
\caption{Analysis of the $f_{\rm X}-f_{\rm UV}$ relation in redshift bins. \emph{Top panel}: slope $\gamma$ as a function of redshift. The slope stays around the reference value of $0.6$ for the entire redshift range explored. 
\emph{Bottom panel}: dispersion as a function of redshift. The dispersion stays around the median value $\sim$0.25 dex, showing a decreasing trend with redshift. This is easily explained by the fact that, at the highest redshifts, only pointed, dedicated observations, unaffected by the observational issues characterising serendipitous observations, are available. Both intrinsic dispersion (output of the \emph{emcee} regression analysis ) and observed one are shown.}
    \label{fig:recap_zbins}
  \end{center}
\end{figure*}

Figs.~\ref{fig:zbins}, \ref{fig:lo_lx}, \ref{fig:recap_zbins} and Tab.~\ref{tab3} show the main results of our analysis. First, the slope of the $f_{\rm X}-f_{\rm UV}$ relation does not evolve with redshift, with an average value around $\langle\gamma\rangle=0.58\pm0.06$ up to $z\simeq4.5$. Second, the mean intrinsic dispersion of the relation has an average around $\delta \simeq 0.20$, with a decreasing trend with redshift.

Fig.~\ref{fig:recap_zbins} summarises the results on the slope and intrinsic dispersion of the $f_{\rm X}-f_{\rm UV}$ relation as a function of the median redshift for the 17 bins. The regression analysis has been performed on the final selected sample ($\Gamma=1.7$--2.8, $k  \, \delta_{obs}=0.6$). As for the slope, it stays around the reference value ($\gamma=0.6$, orange line) within 1$\sigma$ for most of the bins (13/17) and within 2$\sigma$ for all but one point ($\langle z \rangle=1.99$), which is however consistent within 3$\sigma$ with the expected value. This result not only implies that the same physical process empowering the hot corona for the X-ray emission has to be present since when quasars were younger objects (at least at $z \sim 4$), but also that we can safely rely on the $f_{\rm X}-f_{\rm UV}$ relation for measuring cosmological distances and build the Hubble diagram for quasars, the tool that allows us to probe for the first time the space of cosmological parameters in the redshift range $z \sim 1.4$--5.

One of the most important results that we obtained is shown in the bottom panel of Fig.~\ref{fig:recap_zbins}, where the violet data points show the observed dispersion for the $\Delta \rm{log} z = 0.06$ redshift bins, while the light blue ones show the corresponding intrinsic dispersion, output of the \emph{emcee} regression analysis (the same legend of the upper panel applies here for the grey data points). While moving towards higher redshifts, the dispersion drops constantly to smaller values compared to the average value of lower redshift bins.
Given the limited number of sources in the high redshift bins, and the presence of a couple of points with large error bars in $\rm{L_{\rm UV}}$ that ease the tightness in the relation, we conservatively take as a reference the observed dispersion. There is a clear indication of a decrease in the dispersion at high redshift: the observed dispersion (in violet) in the last three bins ranges around 0.06--0.20 dex (Table~\ref{tab3}). Similar results are obtained for $\Delta \rm{log} z = 0.05$ and $0.07$, with the remarkable exception of the second $\Delta \rm{log} z = 0.05$ bin for which the relation has significantly lower dispersion ($\delta_{obs}=0.10$ dex) with respect to the other bin sizes.
This behaviour is the result of the inclusion of distant sources that have been detected with dedicated observation, as opposed to most closer sources, whose presence in the catalogue is based on serendipitous observations. This is shown in Fig. \ref{fig:zbins}, where the inset histograms in the right corner of each z-bin subplot show the distribution of the off-axis angle $\Theta$ for the sources in the bin. For higher $z$, the contribution of sources with $\Theta < 5'$ becomes progressively more important. As proved by a sub-sample of 18 sources at $z \sim 3$ observed with {\it XMM-Newton} \citep{RisalitiLusso2019, Nardini2019, lusso2020}, performing X-ray dedicated observations is crucial to reducing the dispersion associated with calibration issues in the X-ray band. For this sub-sample of 18 sources only, our group performed a spectral analysis both in both the UV and the X-ray band, obtaining a final dispersion of $\delta=0.12$, to be compared to the $\delta=0.24$ of the entire, photometric sample \citep{RisalitiLusso2019}.
The very low background level of the spectra in the CSC 2.0, where pointed \emph{Chandra} observations were performed, allowed us to improve the fits of the relation well beyond $z\simeq 3$.
In fact, the uncertainties in the X-ray measures are consistent with those in the UV (obtained through a fitting of the DR7 spectra and through an interpolation of the photometric SED for the remaining quasars). Quasars from the \emph{XMM-Newton} sample that were drawn from the catalogue of serendipitous sources do not go beyond redshift $z \sim 3.3$. Measures beyond this redshift are mostly from pointed observations (see \citealt{lusso2020} for further details) available in the literature, mostly  \emph{Chandra} observations covering a redshift range $z=4.01$-$7.08$. These have been studied by our group with the aim of examining the relation at the highest redshifts \citep{Salvestrini2019}. For the latter sample, we found a slope $\gamma =0.53 \pm 0.11$, flatter than those in our higher redshift bins, but fully consistent (within 1 $\sigma$) with our results. In the present study, leveraging on \emph{Chandra}'s capabilities, we are therefore able to significantly extend the use of catalogue sources with respect to the \emph{XMM-Newton} sample, up to $z \sim 4.5$.

This result is crucial for the cosmological application of quasars through the $f_{\rm X}-f_{\rm UV}$ relation, that is for establishing whether or not we can consider quasars as standard candles. In principle, with such a small dispersion, we can measure cosmological distances with an even better precision than with supernovae ($\delta_{\rm{SNe}} \sim 0.06$), and at redshifts that supernovae could never probe in similar numbers ($z>2$). Even when the dispersion is higher, if we consider for example the \emph{observed} dispersion in the last three bins ($\langle \delta \rangle \sim 0.15$), the amount of cosmological information provided by one supernova Ia can be achieved with $(0.15/0.06)^{2} \sim 7$ quasars, with the important difference that no supernova Ia has ever been spectroscopically confirmed at redshift higher than 2.26 \citep{Scolnic2018}, whereas hundreds of thousands of quasars are instead being discovered and observed by extragalactic all-sky survey in the last years, and even the most distant objects have already been targeted with dedicated X-ray observations up to redshift $z\sim7.5$ \citep{Banados2018}.
While the decrease in the dispersion with respect to previous works is fully appreciable in the analysis of the $f_{\rm X}-f_{\rm UV}$ relation, especially in the high-redshift bins, it appears to be diluted when we analyse the relation between luminosities (Fig.~\ref{fig:lo_lx}). As a result, the dispersion is comparable to the one in the final SDSS-XMM sample ($\delta \sim 0.24$).
The similarity in the dispersion found for the \emph{Chandra} and the \emph{XMM-Newton} samples, even if for the first one a complete spectroscopical analysis was performed while for the second one the flux at 2 keV was inferred from photometry, is not surprising if we consider that the \emph{Chandra} sample is, on average, fainter than the \emph{XMM} one\footnote{ For the final selection, the distribution of $\rm{log} f_{\rm{UV}}$ has a mean (median) value of $-27.72 \pm 0.57$ ($-27.72_{-0.35}^{+0.36}$) and $-27.48 \pm 0.45$ ($-27.50_{-0.28}^{+0.29}$) for the \emph{Chandra} and \emph{XMM-Newton} samples, respectively. Similarly, the $\rm{log} f_{\rm{X}}$ distribution has a mean (median) value of $-31.60\pm 0.46$ ($-31.59_{0.35}^{+0.30}$) and $-31.36 \pm 0.44$ ($-31.41_{-0.26}^{+0.29}$) for the \emph{Chandra} and \emph{XMM-Newton} samples, respectively. The values listed as confidence intervals for the median are the 25$^{\rm{th}}$ and 75$^{\rm{th}}$ percentiles.}.
A thorough comparison of these samples can be found in \cite{lusso2020}, where the overlap of $\sim 200$ sources at $z<3$ between the two is taken into account.

\subsection{Contribution of X-ray variability to the dispersion}
\begin{figure*}
\begin{center}
	\includegraphics[width=2.\columnwidth]{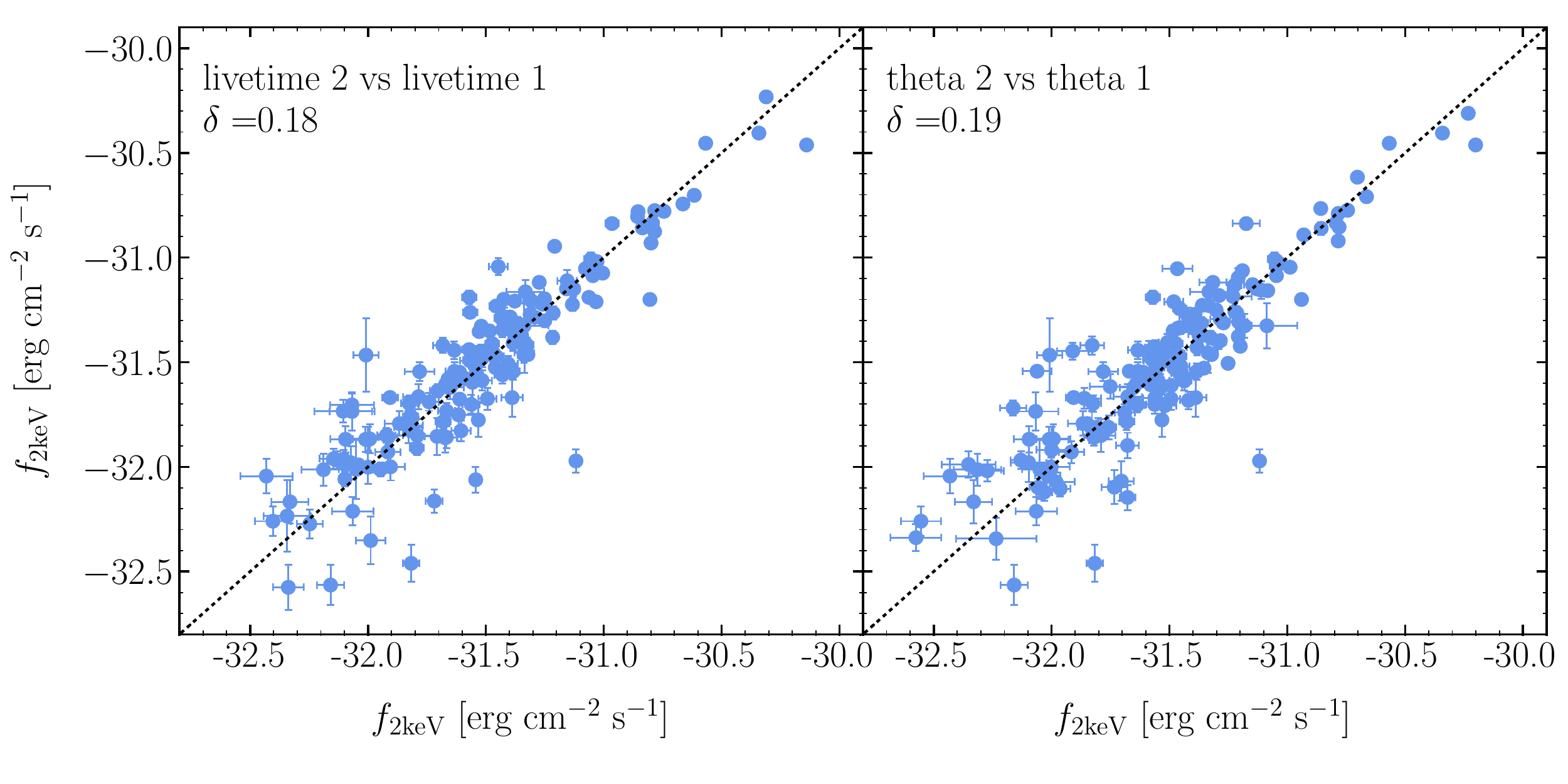}
\caption{Comparison of the rest-frame 2-keV flux as inferred from the longest versus the second longest observation (\emph{left panel}), and from the observation characterised by the smallest versus the second smallest off-axis angle $\Theta$. The dispersions with respect to the 1:1 relation are a combination of the contribution due to the variability in the X-ray band and to observational issues, especially to flux calibration, related to different off-axis angle of the source in different observations.}
    \label{fig:mult_long_small}
    \end{center}
\end{figure*}

\begin{figure}
\begin{center}
	\includegraphics[width=1.\columnwidth]{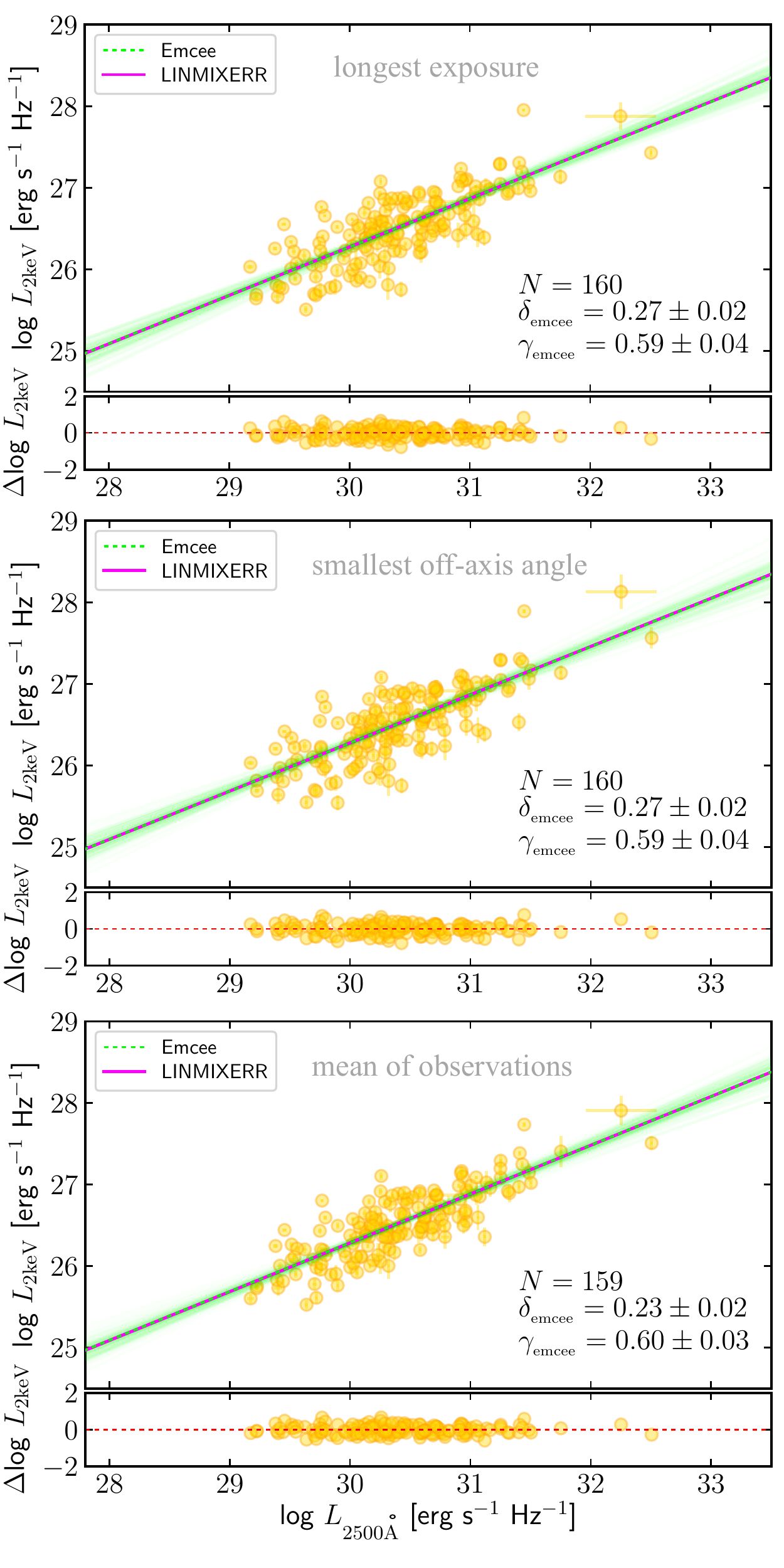}
\caption{$L_{\rm X}-L_{\rm UV}$ relation for the sub-sample of 160 sources with multiple observations. The X-ray luminosity has been inferred for each source from the longest observation (top panel), from the one with the smallest off-axis angle (mid panel) and from an arithmetic mean of all the fluxes available, that is of all the observations surviving the quality filters that pertain to the single source.}
    \label{fig:rel_mult}
    \end{center}
\end{figure}

We have reduced the dispersion in the $f_{\rm X}-f_{\rm UV}$ relation by using spectroscopic data for the X-ray band, especially in the high redshift bins, where all the observations are dedicated. However, we still have to account for many other contributors, among which there are some we can address (e.g. variability in the X-rays and in the UV band) and some instead we can only acknowledge (e.g. radiations emitted by accretion disc and hot corona are intrinsically delayed, preventing an actual simultaneous monitoring of the two structures).
As already mentioned, the CSC 2.0 makes available all the observations carried out for the sources in the catalogue. After having applied all of our filters in both the UV and X-ray band, we therefore have a subset of sources for which multiple  observations ($N_{\rm obs}\geq2$) are available (164/958, 17\% of the sample). 
By analysing the $\Lx-\Lo$ relation for the sources with multiple observations, we can give an estimate of the contribution of X-ray variability to the dispersion in our sample. 
The best value for the 2 keV flux can be chosen using two different criteria: we can opt for the observation with the longest exposure time or the one in which the source has the smallest off-axis angle $\Theta$. In both cases we are looking for an optimisation of the source against the background flux, and for the most accurate and precise measure of the X-ray flux. Both, however, will be necessarily dependent on the flux level at the moment of the observation, namely on the variability of the source.
We compared the 2-keV flux for the observations with the longest and second longest exposure times  (Fig.~\ref{fig:mult_long_small}, left panel), and those for the smallest and second smallest off-axis angle (right panel). The dispersion in the relation with respect to the bisector is, in both cases, the combination of the contribution due to the variability in the X-ray band with the contribution due to the off-axis angle of the source, that is a flux-calibration issue in the X-ray band.
To minimise these contributions, we performed an arithmetic mean of the 2-keV fluxes for all the observations of each source. The result is shown in Fig.~\ref{fig:rel_mult}, where we compare the relation using the observation with the longest exposure (top panel), the smallest off-axis angle $\Theta$ (mid panel) and the arithmetic mean of all the observations available. The dispersion for the sub-sample of sources with multiple observations is characterised by an overall higher dispersion ($\delta \sim 0.27$ dex) if compared with the total sample ($\delta \sim 0.24$ dex). In the case of the mean, however, we observe a significant decrease in the dispersion ($\delta \sim 0.23$ dex), from which we can give an estimate of the contribution of the variability of
\begin{equation}
\sqrt{0.27^{2} - 0.23^{2}} \sim 0.14 \, \mathrm{dex} \,.
\end{equation}
This result is of the same order of magnitude of what was found from the comparison in Fig.~\ref{fig:mult_long_small}, and it is in agreement with \cite{LussoRisaliti2016}. 

\subsection{Additional contributions to the dispersion}
Another contribution to the dispersion is associated with the variability in the UV band, even if to a much lesser extent \citep[$\sim$0.05 mag;][]{sesar2007,Kozlowski2010,ai2010}. Light-curve fitting is the ideal tool to quantify variability. Multiple photometric observations spanning several years and calibrated to better than 0.02 mag are required to produce well-sampled light curves for each object. The relatively low number of optical/UV observations (two on average, performed during the SDSS and BOSS campaigns) for a small number of objects in our final cleaned sample ($<20\%$ have more than one observation, SDSS or BOSS, and only 4\% have two sets of $ugriz$ magnitudes from SDSS and BOSS) and the sparse sampling do not allow a reliable quantification of variability using this methodology in our sample.
Low-order statistics (e.g. rms scatter) is a viable option but the number of objects in our final sample for which this analysis can be performed is so small that any number would be affected by high uncertainties, much higher than the observed variability amplitude.
Some additional artificial variability introduced by the non-simultaneity of the X-ray and UV data could be present, although it does not represent a significant contribution to the dispersion (see Section~5 in \citealt{LussoRisaliti2016}).

An additional important element to be taken into account is the orientation of the accretion disc, which causes its intrinsic emission to be scaled by a function of the inclination angle.
We are planning to apply a correction for the disc inclination with respect to the line of sight by making use of the rest-frame equivalent width of the [\oiii] emission line, EW[\oiii], for the sources for which this line is available in SDSS ($z<0.8$). This quantity has been proposed by our group as an orientation indicator of the accretion disc \citep{Risaliti2011,Bisogni2017,Bisogni2019}.
We will further analyse this aspect in a future work.

\subsection{Simulating the effect of observational contaminants to the dispersion}

To verify whether the decrease in the dispersion examined in this study is reasonable, we performed the following exercise: starting from a simulated sample, we progressively added the observational effects that we have removed from the real one.
We started by bootstrapping a distribution of UV fluxes from the data-sets composed by 2,332 SDSS DR14 and 288 COSMOS sources (2,620 quasars in total) with a measured flux limit at X-rays. 
As discussed in Section 2, the above sample was selected to be homogeneous in the optical/UV, where observational biases and contaminants are minimised.
To a reasonable extent, the sample contains only sources free of - or very poorly affected by - dust reddening and host galaxy contamination, which can be used as a suitable pool for the simulated fluxes.
Fainter objects are preferentially mapped, in the X-ray band, by the deeper Chandra COSMOS Legacy survey, with a flux limit per observation and a flux error that are smaller than those of CSC 2.0 observations. In order to shape a realistic sample, we performed 2,000 extractions of UV fluxes (similar in size to the original sample) and, contextually, extracted errors for the UV and X-ray measurements, and a flux limit value in the X-ray band.
Assuming a perfect, that is dispersion-free, $\Lx-\Lo$ relation, we inferred fluxes at 2 keV for all the sources.

We first simulated the contribution of X-ray variability to the dispersion in the $\Lx-\Lo$ as follows.
We assumed X-ray variability for a source to be represented by a log-normal distribution centred on its average flux value, corresponding, in our case, to the `expected' $\Lx$ inferred directly from the relation  with a standard deviation $\sigma$ representing the characteristic fractional variability of the sources \citep{Maughan2019}. 
The effect of variability was simulated by extracting a random value from the log-normal distribution for each source to reproduce the observed $\Lx-\Lo$ relation.

AGN variability is found to increase with rest-frame timescale \citep{Paolillo2004,Vagnetti2011}, reaching up to 50\% on timescales of decades \citep{Middei2017}, but it is also found to depend on the source luminosity \citep{Vagnetti2016}. The characteristic fractional variability $\sigma$ was then chosen depending on the X-ray luminosity of the source. 
Following \cite{Vagnetti2016}, we assigned a range of fractional variability increasing with 0.5--4.5 keV luminosity\footnote{The luminosity in the 0.5--4.5 keV band was computed from the luminosity at 2 keV by adopting a photon index of $\Gamma=1.7$ \citep{Vagnetti2016}.} of $L <10^{44}$, $10^{44}<L<10^{45}$, $L>10^{45}$ erg s$^{-1}$. The $\sigma$ to be associated with a source, was computed as the mean of 1,000 extractions from a uniform distribution of a fractional variability ranging from 0 to a maximum value that depended on the luminosity of the source (0.45 for the faintest, 0.26 for the intermediate, and 0.14 for the brightest sources).   
An extraction from the log-normal distribution for each source provided the `observed' $\Lx$ value. 

A fit to the simulated data, when a single value was extracted from the log-normal distribution, matched the dispersion of 0.14 dex due to variability in the real sample and gave a slope $\gamma=(0.65\pm0.01)$. When the mean of 1,000 extractions was used instead, the same dispersion dropped to $\sim$0.12 dex.

The inclusion of sources on an emission level higher than their average is preferentially located in - even if not limited to - the low-luminosity end of the relation,
and causes a variation in the overall slope of the relation.
The flux limits associated to the simulated sample were used to exclude the sources with flux at 2 keV close to the observation flux limit (Eddington bias). This step  did not significantly modify the slope observed in the relation and reduced only slightly the amount of dispersion ($\delta=0.13$ dex vs $0.14$ dex). 

Finally, to account for the unlikely presence of heavy obscuration ($N_{\rm{H}}>10^{22}$cm$^{-2}$) in the sample, we performed a simulation to roughly quantify this effect. It is important to note that, in the redshift range covered by our sample, we are observing the rest-frame hard band which is less affected by the presence of absorption.
Specifically, we multiplied the luminosities for the exponential term $e^{-N_{\rm{H}} \, \sigma_{E}}$, where the photoionisation cross-section $\sigma_{E}$ has been computed at 2 keV following \cite{Verner1996}.
We then considered $N_{\rm{H}}$ ranging from $0.5 \times 10^{21}$ to $0.5 \times 10^{23}$ cm$^{-2}$, finding a slope of the relation steadily around $\sim$0.6 and an intercept that decreases progressively from $6.8\pm0.2$ ($N_{\rm{H}}=0.5 \times 10^{21}$ cm$^{-2}$) to $6.7\pm0.2$ for($N_{\rm{H}}=10^{22}$ cm$^{-2}$) and $6.4\pm0.2$ for the most heavily obscured cases ($N_{\rm{H}}=0.5 \times 10^{23}$ cm$^{-2}$), while the intercept for the real sample
is $8.5 \pm 0.4$. 
For column densities larger than that, $\gamma$ starts decreasing sensibly. Even if some obscuration in the X-ray band is left in the sample after the selection in terms of the photon index $\Gamma$, a comparison between slope and intercept in the case of the real data and those for the simulated sample sets $N_{\rm{H}} \sim 10^{22}$ cm$^{-2}$ as an upper-limit estimate of the column density for the sources in the original sample. This simulation is meant to be highly conservative and overestimates the number of absorbed sources.
A similar effect is found if we assume absorption in the optical band: lower optical luminosities 
for larger extinction result in an increasing intercept of the relation.

\section{Conclusions} \label{sec:conclusions}

We presented an analysis of the $\Lx-\Lo$ relation on a sample of 3,430 observations (2,332 sources) from the latest release of the \emph{Chandra} Source Catalog (CSC 2.0), and of 273 sources from the \emph{Chandra} COSMOS Legacy survey. The aim of this work was to employ the largest spectroscopic X-ray sample available up to now with \emph{Chandra} to decrease the dispersion in the relation, verifying its non-evolution with redshift and examining the X-ray variability contribution thanks to the multiple observations available. We pre-selected the parent sample to include only the sources representative of the intrinsic relation, that is type-1, radio-quiet, non-BAL, non-dust-reddened quasars. We then filtered out all the observations that did not meet the required quality criteria in the X-ray band: we excluded possibly absorbed sources characterised by a photon index $\Gamma<1.7$, and observations affected by the Eddington bias. 
The analysis of the relation delivers the following results:
\begin{itemize}
\item[1.] The X-ray-to-UV flux relation analysed in small redshift bins - small enough to make the difference in the luminosity distance for sources in the same bin negligible - does not show any statistically significant evolution with redshift. The slope stays around the expected value of 0.6 within the entire  redshift range probed by our sample ($z \sim 0.5$--4.5).
\item[2.] The dispersion in the relation between fluxes is in agreement with what found in previous works ($\langle \delta \rangle \sim 0.24$ dex) up to $z \sim 3$. It strongly decreases in the last redshift bins ($\langle \delta_{obs} \rangle \sim 0.15$ dex). 
This behaviour is explained by two facts: {\it (a)} at the highest redshifts, the majority of the objects have dedicated, pointed X-ray observations for which calibration issues are minimised; {\it (b)} the decrease is also ascribable to the use of spectroscopic data in the X-ray band  and to the very low level of background of \emph{Chandra} observations, 
\item[3.] Over the entire clean sample, the relation between the X-ray and UV/optical luminosities has a dispersion of $\sim$0.24 dex, comparable to the one found by previous works with samples of similar size.
\item[4.] The analysis of the sub-sample of sources with multiple observations available shows that the mean 2-keV flux, obtained from all the observations that have survived the quality filters, minimises the dispersion in the $\Lx-\Lo$ relation with respect to the choice of the single `best' observation, in terms of both longest exposure time and smallest off-axis angle. Performing a mean of the observations corresponds to removing part of the dispersion, mostly due to X-ray variability, but also to possible issues in the flux calibration related to the off-axis angle of the source. We estimated a contribution of the X-ray variability or calibration issues of the order of $\sim$0.14 dex, in agreement with previous results.
\item[5.] The analysis of observational contaminants on a simulated quasar sample confirms the amount of dispersion in the relation ascribable to X-ray variability ($\delta \sim 0.14$ dex) and allows an estimate of the upper limit for the average column density in the sample ($N_{\rm{H}} \leq 10^{22}$ cm$^{-2}$).
\end{itemize}
Points 1. and 2. have major implications. First of all the interplay between accretion disc and hot corona over the entire redshift range examined so far has to be universal. The relation is very tight once we exclude any possible contribution to the dispersion that is not intrinsic. Secondly, we can use the relation to robustly infer cosmological distances at every redshift.
These two results further justify the employment of quasars as standardisable candles in cosmology.

\begin{acknowledgements}

      SB acknowledges funding from the INAF Ob.Fun. 1.05.03.01.09 Supporto Arizona - LBT Italia.
      SB was also supported by NASA through the Chandra award no. AR7-18013X issued by the Chandra X-ray Observatory Center, operated by the Smithsonian Astrophysical Observatory for and on behalf of NASA under contract NAS8-03060, and partially by  grant  HST-AR-13240.009.
      
      EL acknowledges the support of grant ID: 45780 Fondazione Cassa di Risparmio Firenze.

      This research has made use of data obtained from the Chandra Source Catalog, provided by the Chandra X-ray Center (CXC) as part of the Chandra Data Archive. 

        Funding for the Sloan Digital Sky Survey IV has been provided by the Alfred P. Sloan Foundation, the U.S. Department of Energy Office of Science, and the Participating Institutions. SDSS-IV acknowledges
        support and resources from the Center for High-Performance Computing at
        the University of Utah. The SDSS web site is www.sdss.org.
        
        SDSS-IV is managed by the Astrophysical Research Consortium for the 
        Participating Institutions of the SDSS Collaboration including the 
        Brazilian Participation Group, the Carnegie Institution for Science, 
        Carnegie Mellon University, the Chilean Participation Group, the French Participation Group, Harvard-Smithsonian Center for Astrophysics, 
        Instituto de Astrof\'isica de Canarias, The Johns Hopkins University, 
        Kavli Institute for the Physics and Mathematics of the Universe (IPMU) / 
        University of Tokyo, the Korean Participation Group, Lawrence Berkeley National Laboratory, 
        Leibniz Institut f\"ur Astrophysik Potsdam (AIP),  
        Max-Planck-Institut f\"ur Astronomie (MPIA Heidelberg), 
        Max-Planck-Institut f\"ur Astrophysik (MPA Garching), 
        Max-Planck-Institut f\"ur Extraterrestrische Physik (MPE), 
        National Astronomical Observatories of China, New Mexico State University, 
        New York University, University of Notre Dame, 
        Observat\'ario Nacional / MCTI, The Ohio State University, 
        Pennsylvania State University, Shanghai Astronomical Observatory, 
        United Kingdom Participation Group,
        Universidad Nacional Aut\'onoma de M\'exico, University of Arizona, 
        University of Colorado Boulder, University of Oxford, University of Portsmouth, 
        University of Utah, University of Virginia, University of Washington, University of Wisconsin, 
        Vanderbilt University, and Yale University.
        
    This work made use of matplotlib, a Python library for publication quality graphics \citep{Hunter2007}, and of the software for the analysis and manipulation of catalogues and tables TOPCAT \citep{Taylor2005}.
\end{acknowledgements}

%
%
\bibliographystyle{aasjournal} 
\balance
\bibliography{CSC2}


\begin{appendix} 

\section{UV/optical/NIR SED analysis} \label{sec:appendix1}

In this Appendix we give a detailed description of the analysis in the UV/optical/NIR bands for the two sub-samples.
\nobalance

\begin{table}[ht]
\footnotesize
\caption{Instruments and filters for UV/optical/NIR photometry for the SDSS-CSC2.0 and the Chandra COSMOS Legacy samples.}
\begin{center}
\renewcommand{\arraystretch}{1.0}
\begin{tabular*}{1.0\linewidth}{@{\extracolsep{\fill}}l l c }
\hline
Band          & Instrument       &  $\lambda$  \\
                 &                         &          (\AA)           \\
\hline 

  &  \,\,\,\,\,\,\,\,\,\,\,\,\,\,\,\,\,{\bfseries SDSS - CSC 2.0} &    \\

\hline
\textit{FUV}  & \textit{GALEX}&     $1539$        \\
\textit{NUV}  &  \,\,\,\,\, -     &     $2316$       \\
    \textit{u}         &   SDSS &   $3557$       \\
    \textit{g}          &  \,\,\,\,\, - &   $4825$        \\
   \textit{r}          &  \,\,\,\,\, - &   $6261$       \\
  \textit{i}          &  \,\,\,\,\, - &   $7672$      \\
   \textit{z}           &  \,\,\,\,\, - &   $9097$       \\
\textit{Y}  &UKIDSS &     $10305$         \\
\textit{J}  &  \,\,\,\,\, - &     $12483$        \\
\textit{H}  &  \,\,\,\,\, - &     $16313$      \\
\textit{K}  &  \,\,\,\,\, - &     $22010$       \\
\textit{J}  &2MASS &     $12350$         \\
\textit{H}  &  \,\,\,\,\, -  &     $16620$        \\
\textit{K}  &  \,\,\,\,\, -  &     $21590$        \\

\textit{W$_{1}$}  &\textit{WISE} &     $34000$       \\
\textit{W$_{2}$}  &  \,\,\,\,\, -    &     $46000$         \\
\textit{W$_{3}$}  &  \,\,\,\,\, -    &     $120000$        \\
\textit{W$_{4}$}  &  \,\,\,\,\, -    &     $220000$        \\

       ch1   & \textit{Spitzer} (IRAC)  &     $36000$           \\
       ch2   &   \,\,\,\,\, -                  &     $45000$          \\
      ch3  &    \,\,\,\,\, -                  &     $58000$          \\
       ch4  &   \,\,\,\,\, -                   &     $80000$          \\
\hline  

  &   \,\,\,\,\,{\bfseries Chandra COSMOS Legacy} &    \\

\hline

\textit{NUV}  & \textit{GALEX}&     $2316$          \\
\textit{u}  & CFHT(MegaCam) &     $3881.58$          \\
\textit{B}  &SUBARU (Suprime-Cam)&     $4478$         \\
\textit{V}  & \,\,\,\,\, - &     $5493$        \\
\textit{r}  & \,\,\,\,\, -&     $6315.$        \\
\textit{i}  &  \,\,\,\,\, -&     $7683.9$     \\
               &  CFHT (WIRCam)   &    $7617.66$       \\
                &   SDSS &   $7672.0$                   \\
\textit{z$^{+}$}  &SUBARU (Suprime-Cam) &     $9021.6$      \\
\textit{Y-HSC}  & SUBARU (HyperSuprimeCam) &     $9779.93$         \\
\textit{Y}       &VISTA (VIRCAM)&        $10214.2$       \\
\textit{J}  & \,\,\,\,\, -&     $12534.6$          \\
\textit{H}  & \,\,\,\,\, -&     $16453.4$           \\
\textit{K$_{S}$} &  \,\,\,\,\, - &     $21539.9$    \\
                         &   CFHT (WIRCam)       &    $21480.2$   \\
ch1       & \textit{Spitzer} (IRAC)  & $35550$    \\
                                &   \,\,\,\,\, - &  $35634.3$  \\
       ch2   &  \,\,\,\,\, -  &     $45110.1$            \\
      ch3  &  \,\,\,\,\, -  &     $57593.4$            \\
       ch4  &   \,\,\,\,\, - &     $79594.9$          \\
\hline

\end{tabular*}
\label{tab4}
\end{center}
\end{table}

\noindent {\it {\bfseries SDSS-CSC2.0:}} The photometry available allowed us to build an interpolated SED for each individual object, from which we inferred the rest-frame flux at $2500$\,\AA, proxy for the accretion disc emission. As discussed below, we are interested in estimating the continuum up to 1 $\mu$m rest-frame. For sources at $z\sim2$ and above, this wavelength falls into the mid-infrared and near-infrared bands.
We therefore used all the magnitudes provided by \cite{Paris2018DR14}, that is \emph{GALEX}, SDSS, UKIDSS, 2MASS, \emph{WISE} and \emph{Spitzer}, whose wavelengths are listed in Tab. \ref{tab3}.
We first corrected the fluxes densities in each band for Galactic extinction, using the $E(B-V)$ values estimated from the \cite{Schlegel1998} maps, as listed in the NASA/IPAC Infrared Science Archive\footnote{\url{https://irsa.ipac.caltech.edu/applications/DUST/}} (IRSA), and the reddening law by \cite{Fitzpatrick1999} with $R_{V}=3.1$.
We then computed the rest-frame fluxes according to the redshift listed in \cite{Paris2018DR14} and performed a linear interpolation to retrieve the flux (and luminosity) value at $2500$\AA\ (a higher-order polynomial could introduce spurious features and lead to higher inaccuracy in the UV flux estimation). 
When the combination of source redshift and data available did not provide the wavelength range to cover the rest-frame 2500\,\AA, the flux was extrapolated from the last two photometric points available.
Continuum fluxes estimated in this way can be contaminated by the presence of the broad emission lines characterising type-1 quasars spectra. Depending on the source redshift, broad emission lines are found in different bands and therefore affect different observed magnitudes. In the case of SDSS-CSC2.0 data, we were able to compute a correction taking advantage of the spectral fitting performed by \cite{Shen2011} on DR7 quasars. \cite{Shen2011} take into account continuum, broad and narrow emission lines and Fe\,\textsc{ii} emission, hence providing an accurate measure of the $2500$-\AA\ continuum flux. Comparing our photometric measures with the spectral ones by \cite{Shen2011}, we inferred a correction as a function of redshift, that could be applied to all the SDSS-CSC2.0 photometric $2500$-\AA\ flux estimates.

\begin{figure}
	\includegraphics[width=1.0\columnwidth]{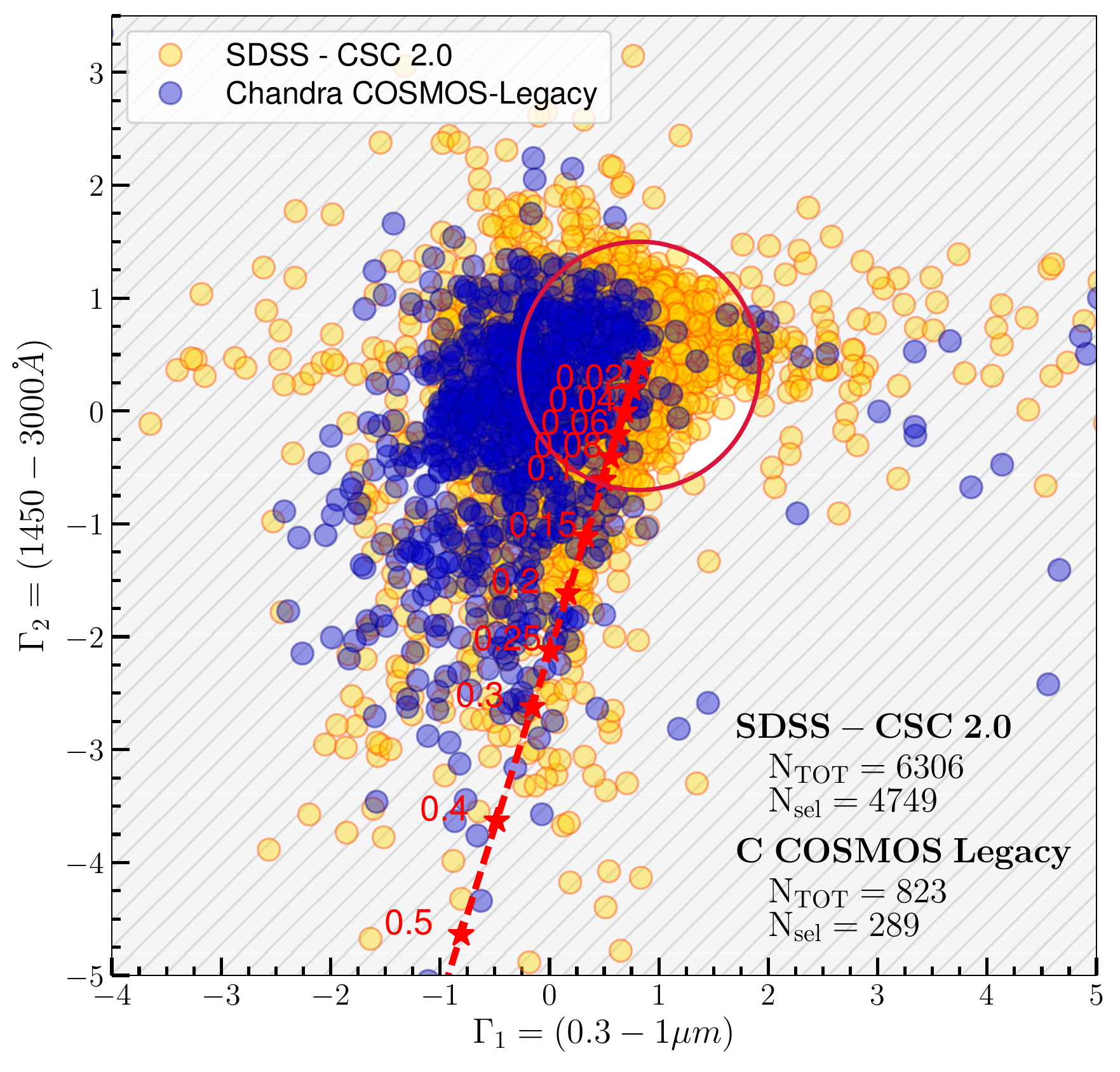}
\caption{Distribution of the quasar sample (excluding radio-loud and BAL sources) in a $\Gamma_{1}$--$\Gamma_{2}$ plane, where $\Gamma_{1}$ and $\Gamma_{2}$ are the slopes of a power law in the $\log\nu$–($\log\nu L_{\nu}$) plane, at 0.3--1\,$\mu$m and 1450-–3000\,\AA, respectively. The red stars represent $\Gamma_{1}$--$\Gamma_{2}$ values for the intrinsic mean SED for quasars of \cite{Richards2006} reddened by the presence of dust (extinction law of \cite{Prevot1984}) with increasing E(B-V), ranging from $0$ to $0.5$. We selected the sources inside the red circle of centre $E(B-V)=0$ and radius $\sim 1$ (corresponding to $E(B-V)=0.1$), characterised by the minimum extinction. Other possible explanations to the deviation of points from the locus of blue quasars are, for example, the presence of host-galaxy contamination and, trivially, wrong photometric measurements.}
    \label{fig:reddening_selection}
\end{figure}

\begin{figure*}
\begin{center}
\includegraphics[width=1.5\columnwidth]{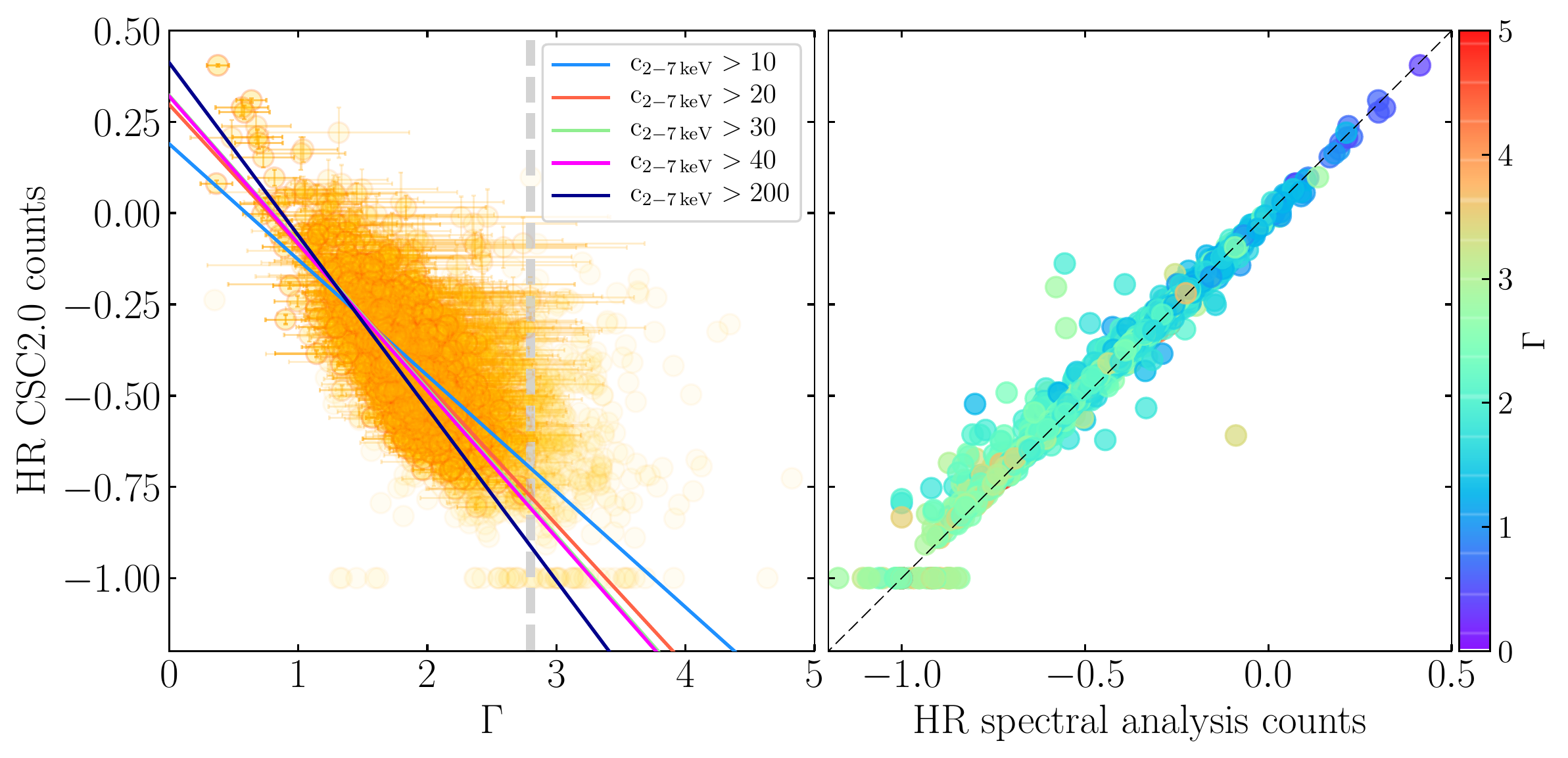}
\caption{Left. Distribution of the HR computed from the catalogued counts as a function of the photon index output of our spectral analysis. Right. Comparison between the HR computed from the catalogued counts and the one resulting from our spectral analysis. Points are coloured as a function of the source photon index (right bar).}
    \label{fig:hardness_ratio}
\end{center}
\end{figure*}

Rest-frame flux (and luminosity) at 1450\,\AA, 3000\,\AA\ and 1 $\mu$m were also computed. Through a simple linear least-square regression the spectral slopes of the power law in log($\nu$)--log($\nu L_{\nu}$) describing the SED ($\Gamma_{1}$, between 0.3 and 1 $\mu$m, and $\Gamma_{2}$, between 1450\,\AA\ and 3000\,\AA) were estimated, and used to estimate the effect by dust reddening and host-galaxy contamination, following the same criteria as in \cite{LussoRisaliti2016}. Specifically, we excluded from the sample the sources in the $\Gamma_{1}$--$\Gamma_{2}$ plane not included in the circle of radius $\sim$1 (Fig. \ref{fig:reddening_selection}), corresponding to $E(B-V) \sim 0.1$, and centred in $E(B-V)=0$, that is the value of the slopes for which the UV spectrum is completely unaffected by dust reddening or host contamination ($\Gamma_{1}=0.82$ and $\Gamma_{2}=0.4$), based on the intrinsic mean SED for quasars of \cite{Richards2006}. For DR7 sources, we used the $2500$-\AA\ flux made available by \cite{Shen2011} through spectral fitting. Since \cite{Shen2011} do not provide errors on this quantity, we assumed a $2\%$, considering that they list a $3\%$ average error on bolometric luminosity, where also the uncertainty on the bolometric correction is included.

\noindent {\it {\bfseries COSMOS:}} The same analysis was performed, in this case with the photometry by \cite{Laigle2016}, making available deep NIR observations for the COSMOS field from the UltraVISTA and SPLASH surveys.
Ranging from the UV to the infrared bands, as listed in the lower part of Tab. \ref{tab3}, the photometry allowed the determination of the rest-frame 2500-\AA\ flux and the estimation of the SED spectral slopes to exclude sources affected by dust absorption and host-galaxy contamination (Fig. \ref{fig:reddening_selection}), following the same criteria described above for the SDSS-CSC2.0 sample. The 4400-\AA\ flux and luminosity were also computed and used, along with the radio luminosity at 5\,GHz, to infer the radio loudness parameter $R$ and get rid of radio-loud sources, as described in Section \ref{sec:sample}.

\section{Check on hardness ratios} \label{sec:appendix1bis}

We further checked all sources with extreme photon index values derived from the spectral analysis - namely, either very low (flat, $\Gamma<1.4$) or unexpectedly high (steep, $\Gamma>3.0$). Fig. \ref{fig:hardness_ratio}, right panel, shows the comparison between the hardness ratios listed in the CSC 2.0 and those computed from the spectral counts retrieved using XSPEC. In both cases, the hardness ratio is defined as~
\begin{equation}
    HR = \frac{H-S}{H+S} \,,
\end{equation}
where $H$ and $S$ are the net counts in the hard ($2$--7 keV) and soft ($0.5$--2 keV) band respectively. The agreement between the two measurements, the first one based on aperture photometry and the second one on spectroscopic data, is excellent, with the exception of a few outliers. This represents an independent validation of the spectral analysis we carried out.
The observations with a non detection in the hard band, namely $HR=-1.0$ (50 for the CSC 2.0, while only 20 have no counts in the hard band from the spectral analysis) are located in the lower left corner of the panel.

The HR is a coarse measure of the spectral slope, nonetheless we expect to observe a correlation between HR and the photon index.
The left panel of Fig. \ref{fig:hardness_ratio} shows the hardness ratio from the CSC 2.0 as a function of the photon index from the spectral analysis for an increasing number of net counts in the hard band. For sources with a photon index $\Gamma<2.8$, the value chosen as an upper threshold, the CSC 2.0 $HR$ is a good proxy for the photon index, improving with the number of counts in the hard and, consequently, in the soft band. This is an obvious consequence of the fact that points deviating from the bestfit relation in the HR vs $\Gamma$ plane, populating the locus of $\Gamma>2.8$ and HR$<-0.25$, are generally associated with observations with a small number of counts ($<10$ in the hard band).

\section{Choice of selection criteria in X-ray band} \label{sec:appendix2}

\begin{figure*}
\begin{center}
	\includegraphics[width=2.0\columnwidth]{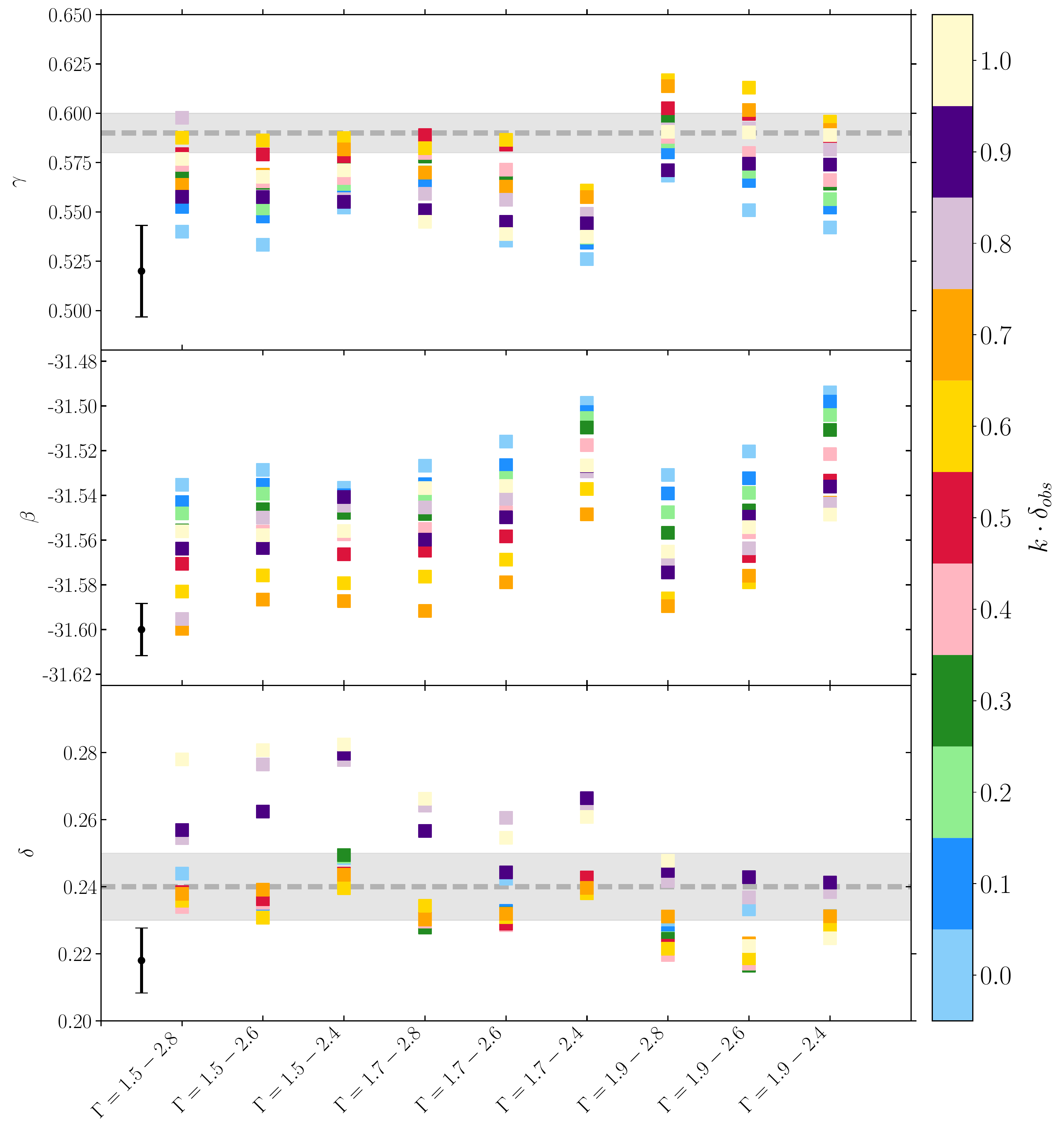}
\caption{Results of the analysis on the $f_{\rm X}-f_{\rm UV}$ relation. Different choices for the photon index $\Gamma$ are on the $x$ axis, while those for $k  \, \delta_{obs}$ are shown as different colours. The squares represent the  mean slope (top panel), intercept (mid panel) and dispersion (bottom panel) for all the redshift bins ($\Delta \mathrm{log}(z)=0.06$) with more than five objects after the chosen cuts in $\Gamma$ and $k  \, \delta_{obs}$ are applied. In black, we show the typical error on the estimates for the three quantities.}
    \label{fig:appendix}
  \end{center}
\end{figure*}

In this work, the parent sample has been pre-selected for the properties concerning the UV band, that is broad-line absorption and dust reddening. 
Here we discuss our choices for the selection in the X-ray band, which led to the final sample of 1,385 CSC 2.0 observations, corresponding to 958 sources, and the 140 COSMOS sources. To decide which combination of selection in photon index $\Gamma$, that is X-ray absorption, and Eddington bias, that is X-ray flux threshold, is a good trade-off between a small observed dispersion and a large enough number of sources, we have analysed the behaviour of the $f_{\rm X}-f_{\rm UV}$ relation for different choices of these quantities. 
Tab.~\ref{tab5} and Fig.~\ref{fig:appendix} summarise the results of the analysis. Fig.~\ref{fig:appendix} shows the slope, intercept and dispersion for the $f_{\rm X}-f_{\rm UV}$ relation  analysed as follows: as explained in Section \ref{sec:analysis}, when dealing with fluxes, we have to divide the sample in redshift bins. 
\balance
The fit to the data in each redshift bin ($\Delta \mathrm{log}(z)=0.06$), gave us a value for slope, intercept and dispersion. Then, for each quantity, we computed a mean of the values for all the redshift bins ($\Delta \mathrm{log}(z)=0.06$) with more than five objects.  Here we are comparing these values for different choices of $\Gamma$ ($x$ axis) and $k  \, \delta_{obs}$ (colour bar)\footnote{In this appendix we are showing the analysis for nine different ranges of $\Gamma$ and eleven different values of $k  \, \delta_{obs}$.}. Values obtained for the $L_{\rm X}-L_{\rm UV}$ relation are plotted in grey as a reference. In Tab.~\ref{tab5} we also list, for each choice of $\Gamma$ and $k  \, \delta_{obs}$ the number of sources that outlived the selection.

We looked at the dispersion and at the slope of the $f_{\rm X}-f_{\rm UV}$ relation: the bottom panel of Fig.~\ref{fig:appendix} shows that, for all the choices of $\Gamma$, the observed dispersion has a minimum for $k  \, \delta_{obs}=0.6$, while it increases for both lower and higher choices of this quantity, meaning not only that the inclusion of sources affected by the Eddington bias (i.e. light blue and pale blue squares) can modify the dispersion in the relation, but also that a too severe selection (i.e. purple, light yellow) implies a drastic decrease in the number of sources and increasing dispersion with respect to best fit.
Moreover, while for a choice of $k  \, \delta_{obs}\sim 0.6$ the slope of the relation stays steadily around $\gamma \sim 0.59$ for every choice of $\Gamma$, for both low and high  $k  \, \delta_{obs}$ values the slope is significantly lower than the reference value of $\sim$0.6 (upper panel).
Finally, exploring the possible choices of the photon index $\Gamma$, we realised that a smaller dispersion is achieved by a more conservative selection at lower values than at higher values; in fact, setting the lower boundary progressively from $\Gamma_{\mathrm{min}}=1.5$ to $\Gamma_{\mathrm{min}}=1.9$ produces a decrease in the dispersion, while this is not generally true going from $\Gamma_{\mathrm{max}}=2.8$ to $\Gamma_{\mathrm{max}}=2.4$. 
Even if the slope of the relation is closer to the reference value and the dispersion is slightly smaller in the case of the most conservative selection $\Gamma=1.9$--2.8, we decided for the compromise $\Gamma=1.7$--2.8 because of the larger number of complying sources, and also for consistency with the cut adopted in the analysis of the SDSS-XMM sample of \cite{RisalitiLusso2019}.
For the sake of completeness, we also plot the intercepts for the different choices of the parameters (middle panel), even if this quantity is not taken into account for the purpose of selection.

\clearpage

\begin{sidewaystable*}

\footnotesize
\caption{Results of the analysis on the $f_{\rm X}-f_{\rm UV}$ relation.}
\begin{center}
\renewcommand{\arraystretch}{0.9}
\begin{tabular}{ C{0.1cm}  C{0.3cm}  C{0.2cm} c c c c c c c c c }
\hline
                                                   &                                           &                                & \multicolumn{9}{c}{$\Gamma$}    \\
                                                   &                                           &                                & $1.5-2.8$ & $1.5-2.6$ &$1.5-2.4$ & $1.7-2.8$ &$1.7-2.6$ &  $1.7-2.4$ & $1.9-2.8$ & $1.7-2.6$ & $1.7-2.4$  \\
\hline
\multirow{44}{*}{\rotatebox{90}{$k\, \delta$ }} & \multicolumn{1}{c|}{\multirow{4}{*}{0.0}} & \multicolumn{1}{c|}{$N$}          &    $         1841          $ &  $            1727           $ &  $           1573           $ &  $           1630          $ &  $          1515            $ &  $         1357          $ &   $          1277          $ &   $         1157           $ &   $          991            $    \\
                                                   & \multicolumn{1}{c|}{}                     & \multicolumn{1}{c|}{$\gamma$}   &    $  0.54    \pm   0.02   $ &  $    0.53    \pm    0.02    $ &  $   0.55    \pm    0.03    $ &  $   0.55    \pm    0.02   $ &  $   0.54    \pm    0.02    $ &  $  0.53   \pm    0.02   $ &   $  0.57    \pm    0.02   $  &  $  0.55    \pm    0.02   $  &  $  0.54    \pm    0.02    $    \\
                                                   & \multicolumn{1}{c|}{}                     & \multicolumn{1}{c|}{$\beta$}    &    $  -31.54  \pm   0.01   $ &  $    -31.53  \pm    0.01    $ &  $   -31.54  \pm    0.01    $ &  $   -31.53  \pm    0.01   $ &  $   -31.52  \pm    0.01    $ &  $  -31.50 \pm    0.01   $ &   $  -31.53  \pm    0.01   $  &  $  -31.52  \pm    0.01   $  &  $  -31.49  \pm    0.01    $    \\
                                                   &     \multicolumn{1}{c|}{}                 & \multicolumn{1}{c|}{$\delta$}   &    $  0.24    \pm   0.01   $ &  $    0.24    \pm    0.01    $ &  $   0.25    \pm    0.01    $ &  $   0.23    \pm    0.01   $ &  $   0.24    \pm    0.01    $ &  $  0.24   \pm    0.01   $ &   $  0.23    \pm    0.01   $  &  $  0.23    \pm    0.01   $  &  $  0.23    \pm    0.01    $    \\           
\cline{4-12} 

                                                   & \multicolumn{1}{c|}{\multirow{4}{*}{0.1}} & \multicolumn{1}{c|}{$N$}          &    $         1778          $ &  $           1672            $ &  $           1533           $ &  $           1571          $ &  $           1464           $ &  $        1322           $ &   $        1226            $ &   $        1114            $ &   $         964             $   \\
                                                   & \multicolumn{1}{c|}{}                     & \multicolumn{1}{c|}{$\gamma$}   &    $  0.55    \pm   0.02   $ &  $    0.55    \pm    0.02    $ &  $   0.56    \pm    0.03    $ &  $   0.56    \pm    0.02   $ &  $   0.56    \pm    0.02    $ &  $  0.53   \pm    0.02   $ &   $  0.58    \pm    0.02   $  &  $  0.57    \pm    0.02   $  &  $  0.55    \pm    0.02    $   \\
                                                   &     \multicolumn{1}{c|}{}                 & \multicolumn{1}{c|}{$\beta$}    &    $  -31.54  \pm   0.01   $ &  $    -31.54  \pm    0.01    $ &  $   -31.54  \pm    0.01    $ &  $   -31.53  \pm    0.01   $ &  $   -31.53  \pm    0.01    $ &  $  -31.50 \pm    0.01   $ &   $  -31.54  \pm    0.01   $  &  $  -31.53  \pm    0.01   $  &  $  -31.50  \pm    0.01    $    \\
                                                   &       \multicolumn{1}{c|}{}               & \multicolumn{1}{c|}{$\delta$}   &    $  0.24    \pm   0.01   $ &  $    0.23    \pm    0.01    $ &  $   0.25    \pm    0.01    $ &  $   0.23    \pm    0.01   $ &  $   0.23    \pm    0.01    $ &  $  0.24   \pm    0.01   $ &   $  0.23    \pm    0.01   $  &  $  0.22    \pm    0.01   $  &  $  0.23    \pm    0.01    $    \\

\cline{4-12}
                                                   & \multicolumn{1}{c|}{\multirow{4}{*}{0.2}} & \multicolumn{1}{c|}{$N$}          &    $         1690          $ &  $           1596            $ &  $          1477            $ &  $          1489           $ &  $         1394             $ &  $        1272           $ &   $          1151          $ &   $         1053           $ &   $         922             $   \\
                                                   &    \multicolumn{1}{c|}{}                  & \multicolumn{1}{c|}{$\gamma$}   &    $  0.56    \pm   0.02   $ &  $    0.55    \pm    0.02    $ &  $   0.56    \pm    0.03    $ &  $   0.57    \pm    0.02   $ &  $   0.56    \pm    0.02    $ &  $  0.54   \pm    0.02   $ &   $  0.59    \pm    0.02   $  &  $  0.57    \pm    0.02   $  &  $  0.56    \pm    0.02    $   \\
                                                   &     \multicolumn{1}{c|}{}                 & \multicolumn{1}{c|}{$\beta$}    &    $  -31.55  \pm   0.01   $ &  $    -31.54  \pm    0.01    $ &  $   -31.54  \pm    0.01    $ &  $   -31.54  \pm    0.01   $ &  $   -31.53  \pm    0.01    $ &  $  -31.51 \pm    0.01   $ &   $  -31.55  \pm    0.01   $  &  $  -31.54  \pm    0.01   $  &  $  -31.50  \pm    0.01    $    \\
                                                   &       \multicolumn{1}{c|}{}               & \multicolumn{1}{c|}{$\delta$}   &    $  0.24    \pm   0.01   $ &  $    0.24    \pm    0.01    $ &  $   0.25    \pm    0.01    $ &  $   0.23    \pm    0.01   $ &  $   0.23    \pm    0.01    $ &  $  0.24   \pm    0.01   $ &   $  0.22    \pm    0.01   $  &  $  0.22    \pm    0.01   $  &  $  0.23    \pm    0.01    $    \\

\cline{4-12}
                                                   & \multicolumn{1}{c|}{\multirow{4}{*}{0.3}} & \multicolumn{1}{c|}{$N$}   &    $          1579         $ &  $           1500            $ &  $          1395            $ &  $         1393            $ &  $         1313             $ &  $       1205            $ &   $        1063            $ &   $         982            $ &   $          867            $   \\
                                                   &     \multicolumn{1}{c|}{}                 & \multicolumn{1}{c|}{$\gamma$}          &    $  0.57    \pm   0.02   $ &  $    0.56    \pm    0.02    $ &  $   0.57    \pm    0.03    $ &  $   0.58    \pm    0.02   $ &  $   0.57    \pm    0.02    $ &  $  0.54   \pm    0.02   $ &   $  0.60    \pm    0.02   $  &  $  0.58    \pm    0.02   $  &  $  0.56    \pm    0.02    $   \\
                                                   &     \multicolumn{1}{c|}{}                 & \multicolumn{1}{c|}{$\beta$}    &    $  -31.56  \pm   0.01   $ &  $    -31.55  \pm    0.01    $ &  $   -31.55  \pm    0.01    $ &  $   -31.55  \pm    0.01   $ &  $   -31.54  \pm    0.01    $ &  $  -31.51 \pm    0.01   $ &   $  -31.56  \pm    0.01   $  &  $  -31.55  \pm    0.01   $  &  $  -31.51  \pm    0.01    $   \\
                                                   &       \multicolumn{1}{c|}{}               & \multicolumn{1}{c|}{$\delta$}   &    $  0.24    \pm   0.01   $ &  $    0.24    \pm    0.01    $ &  $   0.25    \pm    0.01    $ &  $   0.23    \pm    0.01   $ &  $   0.23    \pm    0.01    $ &  $  0.24   \pm    0.01   $ &   $  0.22    \pm    0.01   $  &  $  0.22    \pm    0.01   $  &  $  0.23    \pm    0.01    $    \\

\cline{4-12}
                                                   & \multicolumn{1}{c|}{\multirow{4}{*}{0.4}} & \multicolumn{1}{c|}{$N$}          &    $  1440                 $ &  $           1377            $ &  $          1293            $ &  $          1260           $ &  $          1197            $ &  $        1110           $ &   $         943            $ &   $          879           $ &   $           786           $   \\
                                                   &     \multicolumn{1}{c|}{}                 & \multicolumn{1}{c|}{$\gamma$}   &    $  0.57    \pm   0.02   $ &  $    0.57    \pm    0.02    $ &  $   0.57    \pm    0.03    $ &  $   0.58    \pm    0.02   $ &  $   0.57    \pm    0.02    $ &  $  0.55   \pm    0.02   $ &   $  0.59    \pm    0.02   $  &  $  0.58    \pm    0.02   $  &  $  0.57    \pm    0.02    $   \\
                                                   &     \multicolumn{1}{c|}{}                 & \multicolumn{1}{c|}{$\beta$}    &    $  -31.56  \pm   0.01   $ &  $    -31.56  \pm    0.01    $ &  $   -31.56  \pm    0.01    $ &  $   -31.55  \pm    0.01   $ &  $   -31.55  \pm    0.01    $ &  $  -31.52 \pm    0.01   $ &   $  -31.57  \pm    0.01   $  &  $  -31.56  \pm    0.01   $  &  $  -31.52  \pm    0.01    $   \\
                                                   &       \multicolumn{1}{c|}{}               & \multicolumn{1}{c|}{$\delta$}   &    $  0.23    \pm   0.01   $ &  $    0.24    \pm    0.01    $ &  $   0.24    \pm    0.01    $ &  $   0.23    \pm    0.01   $ &  $   0.23    \pm    0.01    $ &  $  0.24   \pm    0.01   $ &   $  0.22    \pm    0.01   $  &  $  0.22    \pm    0.01   $  &  $  0.23    \pm    0.01    $    \\

\cline{4-12}
                                                   & \multicolumn{1}{c|}{\multirow{4}{*}{0.5}} & \multicolumn{1}{c|}{$N$}          &    $         1289          $ &  $            1237           $ &  $          1168            $ &  $          1120           $ &  $          1068            $ &  $       996             $ &   $          827           $ &   $         775            $ &   $          701            $   \\
                                                   &     \multicolumn{1}{c|}{}                 & \multicolumn{1}{c|}{$\gamma$}   &    $  0.58    \pm   0.02   $ &  $    0.58    \pm    0.02    $ &  $   0.58    \pm    0.03    $ &  $   0.59    \pm    0.02   $ &  $   0.58    \pm    0.02    $ &  $  0.56   \pm    0.02   $ &   $  0.60    \pm    0.02   $  &  $  0.60    \pm    0.02   $  &  $  0.58    \pm    0.02    $   \\
                                                   &     \multicolumn{1}{c|}{}                 & \multicolumn{1}{c|}{$\beta$}    &    $  -31.57  \pm   0.01   $ &  $    -31.56  \pm    0.01    $ &  $   -31.57  \pm    0.01    $ &  $   -31.56  \pm    0.01   $ &  $   -31.56  \pm    0.01    $ &  $  -31.53 \pm    0.01   $ &   $  -31.57  \pm    0.01   $  &  $  -31.57  \pm    0.01   $  &  $  -31.53  \pm    0.01    $   \\
                                                   &       \multicolumn{1}{c|}{}               & \multicolumn{1}{c|}{$\delta$}   &    $  0.24    \pm   0.01   $ &  $    0.24    \pm    0.01    $ &  $   0.24    \pm    0.01    $ &  $   0.23    \pm    0.01   $ &  $   0.23    \pm    0.01    $ &  $  0.24   \pm    0.01   $ &   $  0.22    \pm    0.01   $  &  $  0.22    \pm    0.01   $  &  $  0.23    \pm    0.01    $   \\

\cline{4-12}
                                                   & \multicolumn{1}{c|}{\multirow{4}{*}{0.6}} & \multicolumn{1}{c|}{$N$}          &    $         1133          $ &  $           1088            $ &  $         1025             $ &  $          1098            $ &  $          937             $ &  $        872            $ &   $          720           $ &   $          675           $ &   $          608            $   \\
                                                   &     \multicolumn{1}{c|}{}                 & \multicolumn{1}{c|}{$\gamma$}   &    $  0.59    \pm   0.02   $ &  $    0.59    \pm    0.02    $ &  $   0.59    \pm    0.03    $ &  $   0.58    \pm    0.02   $ &  $   0.59    \pm    0.02    $ &  $  0.56   \pm    0.02   $ &   $  0.62    \pm    0.02   $  &  $  0.61    \pm    0.02   $  &  $  0.60    \pm    0.03    $   \\
                                                   &     \multicolumn{1}{c|}{}                 & \multicolumn{1}{c|}{$\beta$}    &    $  -31.58  \pm   0.01   $ &  $    -31.58  \pm    0.01    $ &  $   -31.58  \pm    0.01    $ &  $   -31.58  \pm    0.01   $ &  $   -31.57  \pm    0.01    $ &  $  -31.54 \pm    0.01   $ &   $  -31.59  \pm    0.01   $  &  $  -31.58  \pm    0.01   $  &  $  -31.55  \pm    0.01    $    \\
                                                   &       \multicolumn{1}{c|}{}               & \multicolumn{1}{c|}{$\delta$}   &    $  0.24    \pm   0.01   $ &  $    0.23    \pm    0.01    $ &  $   0.24    \pm    0.01    $ &  $   0.23    \pm    0.01   $ &  $   0.23    \pm    0.01    $ &  $  0.24   \pm    0.01   $ &   $  0.22    \pm    0.01   $  &  $  0.22    \pm    0.01   $  &  $  0.23    \pm    0.01    $    \\

\cline{4-12}
                                                   & \multicolumn{1}{c|}{\multirow{4}{*}{0.7}} & \multicolumn{1}{c|}{$N$}          &    $          937          $ &  $            904            $ &  $          855             $ &  $          814            $ &  $           781            $ &  $         730           $ &   $          588           $ &   $           555          $ &   $          502            $ \\
                                                   &     \multicolumn{1}{c|}{}                 & \multicolumn{1}{c|}{$\gamma$}   &    $  0.56    \pm   0.03   $ &  $    0.57    \pm    0.03    $ &  $   0.58    \pm    0.03    $ &  $   0.57    \pm    0.03   $ &  $   0.56    \pm    0.03    $ &  $  0.56   \pm    0.03   $ &   $  0.61    \pm    0.03   $  &  $  0.60    \pm    0.03   $  &  $  0.59    \pm    0.03    $   \\
                                                   &     \multicolumn{1}{c|}{}                 & \multicolumn{1}{c|}{$\beta$}    &    $  -31.60  \pm   0.01   $ &  $    -31.59  \pm    0.01    $ &  $   -31.59  \pm    0.01    $ &  $   -31.59  \pm    0.01   $ &  $   -31.58  \pm    0.01    $ &  $  -31.55 \pm    0.01   $ &   $  -31.59  \pm    0.01   $  &  $  -31.58  \pm    0.01   $  &  $  -31.54  \pm    0.01    $   \\
                                                   &       \multicolumn{1}{c|}{}               & \multicolumn{1}{c|}{$\delta$}   &    $  0.24    \pm   0.01   $ &  $    0.24    \pm    0.01    $ &  $   0.24    \pm    0.01    $ &  $   0.23    \pm    0.01   $ &  $   0.23    \pm    0.01    $ &  $  0.24   \pm    0.01   $ &   $  0.23    \pm    0.01   $  &  $  0.22    \pm    0.01   $  &  $  0.23    \pm    0.01    $    \\

\cline{4-12}
                                                   & \multicolumn{1}{c|}{\multirow{4}{*}{0.8}} & \multicolumn{1}{c|}{$N$}          &    $          792          $ &  $            772            $ &  $           734            $ &  $           684           $ &  $           664            $ &  $         624           $ &   $          487           $ &   $           467          $ &   $           425           $   \\
                                                   &     \multicolumn{1}{c|}{}                 & \multicolumn{1}{c|}{$\gamma$}   &    $  0.60    \pm   0.03   $ &  $    0.56    \pm    0.03    $ &  $   0.56    \pm    0.03    $ &  $   0.56    \pm    0.03   $ &  $   0.56    \pm    0.03    $ &  $  0.55   \pm    0.03   $ &   $  0.59    \pm    0.03   $  &  $  0.59    \pm    0.03   $  &  $  0.58    \pm    0.03    $   \\
                                                   &     \multicolumn{1}{c|}{}                 & \multicolumn{1}{c|}{$\beta$}    &    $  -31.60  \pm   0.02   $ &  $    -31.55  \pm    0.01    $ &  $   -31.54  \pm    0.01    $ &  $   -31.55  \pm    0.01   $ &  $   -31.54  \pm    0.01    $ &  $  -31.53 \pm    0.02   $ &   $  -31.57  \pm    0.02   $  &  $  -31.56  \pm    0.02   $  &  $  -31.54  \pm    0.02    $   \\
                                                   &       \multicolumn{1}{c|}{}               & \multicolumn{1}{c|}{$\delta$}   &    $  0.25    \pm   0.01   $ &  $    0.28    \pm    0.01    $ &  $   0.28    \pm    0.01    $ &  $   0.26    \pm    0.01   $ &  $   0.26    \pm    0.01    $ &  $  0.26   \pm    0.01   $ &   $  0.24    \pm    0.01   $  &  $  0.24    \pm    0.01   $  &  $  0.24    \pm    0.01    $    \\

\cline{4-12}
                                                   & \multicolumn{1}{c|}{\multirow{4}{*}{0.9}} & \multicolumn{1}{c|}{$N$}          &    $          617          $ &  $           602             $ &  $           576            $ &  $           536           $ &  $           521            $ &  $         495           $ &   $          392           $ &   $          377           $ &   $         351             $   \\
                                                   &     \multicolumn{1}{c|}{}                 & \multicolumn{1}{c|}{$\gamma$}   &    $  0.56    \pm   0.03   $ &  $    0.56    \pm    0.03    $ &  $   0.56    \pm    0.03    $ &  $   0.55    \pm    0.03   $ &  $   0.54    \pm    0.03    $ &  $  0.54   \pm    0.03   $ &   $  0.57    \pm    0.03   $  &  $  0.57    \pm    0.03   $  &  $  0.57    \pm    0.03    $   \\
                                                   &     \multicolumn{1}{c|}{}                 & \multicolumn{1}{c|}{$\beta$}    &    $  -31.56  \pm   0.02   $ &  $    -31.56  \pm    0.02    $ &  $   -31.54  \pm    0.02    $ &  $   -31.56  \pm    0.02   $ &  $   -31.55  \pm    0.02    $ &  $  -31.53 \pm    0.02   $ &   $  -31.57  \pm    0.02   $  &  $  -31.55  \pm    0.02   $  &  $  -31.54  \pm    0.02    $   \\
                                                   &       \multicolumn{1}{c|}{}               & \multicolumn{1}{c|}{$\delta$}   &    $  0.26    \pm   0.01   $ &  $    0.26    \pm    0.01    $ &  $   0.28    \pm    0.01    $ &  $   0.26    \pm    0.01   $ &  $   0.24    \pm    0.01    $ &  $  0.27   \pm    0.01   $ &   $  0.24    \pm    0.02   $  &  $  0.24    \pm    0.01   $  &  $  0.24    \pm    0.01    $   \\

\cline{4-12}
                                                   & \multicolumn{1}{c|}{\multirow{4}{*}{1.0}} & \multicolumn{1}{c|}{$N$}          &    $           487         $ &  $            476            $ &  $          458             $ &  $           426           $ &  $          415             $ &  $        397            $ &   $          305           $ &   $          294           $ &   $          276            $   \\
                                                   &     \multicolumn{1}{c|}{}                 & \multicolumn{1}{c|}{$\gamma$}   &    $  0.58    \pm   0.04   $ &  $    0.57    \pm    0.04    $ &  $   0.57    \pm    0.04    $ &  $   0.54    \pm    0.04   $ &  $   0.54    \pm    0.04    $ &  $  0.54   \pm    0.04   $ &   $  0.59    \pm    0.04   $  &  $  0.59    \pm    0.04   $  &  $  0.59    \pm    0.04    $   \\
                                                   &     \multicolumn{1}{c|}{}                 & \multicolumn{1}{c|}{$\beta$}    &    $  -31.56  \pm   0.02   $ &  $    -31.56  \pm    0.02    $ &  $   -31.56  \pm    0.02    $ &  $   -31.54  \pm    0.02   $ &  $   -31.54  \pm    0.02    $ &  $  -31.53 \pm    0.02   $ &   $  -31.57  \pm    0.02   $  &  $  -31.55  \pm    0.02   $  &  $  -31.55  \pm    0.02    $   \\
                                                   &       \multicolumn{1}{c|}{}               & \multicolumn{1}{c|}{$\delta$}   &    $  0.28    \pm   0.01   $ &  $    0.28    \pm    0.01    $ &  $   0.28    \pm    0.01    $ &  $   0.27    \pm    0.01   $ &  $   0.25    \pm    0.01    $ &  $  0.26   \pm    0.01   $ &   $  0.25    \pm    0.02   $  &  $  0.22    \pm    0.01   $  &  $  0.22    \pm    0.02    $   \\
\hline     
\end{tabular}
\begin{tablenotes}
{\bfseries Notes.}  Results of the analysis on the $f_{\rm X}-f_{\rm UV}$ relation. We list, for every combination of cuts in $\Gamma$ and $k  \, \delta_{obs}$, the slope, intercept and dispersion obtained as a mean of all those in the redshift bins ($\Delta \mathrm{log}(z)=0.06$) with more than five objects, as long as the number of sources that survived the selection.
\end{tablenotes}
\label{tab5}
\end{center}

\end{sidewaystable*}

\clearpage

\end{appendix}
\end{document}